\documentclass[11pt] {article}
\usepackage[a4paper,margin=1in]{geometry}

\usepackage[T1]{fontenc} %
\usepackage[normalem]{ulem} %

\usepackage[usenames,dvipsnames,svgnames,x11names]{xcolor}

\newcommand{\claim}[1]{\textcolor{black}{\textit{#1}}}

\usepackage{cite}

\usepackage{balance}

\usepackage{listings}

\lstdefinelanguage{XML}
{
basicstyle=\ttfamily\footnotesize,
  morestring=[b]",
  moredelim=[s][\bfseries\color{Maroon}]{<}{\ },
  moredelim=[s][\bfseries\color{Maroon}]{</}{>},
  moredelim=[l][\bfseries\color{Maroon}]{/>},
  moredelim=[l][\bfseries\color{Maroon}]{>},
  morecomment=[s]{<?}{?>},
  morecomment=[s]{<!--}{-->},
  commentstyle=\color{gray},
  stringstyle=\color{blue},
  identifierstyle=\color{red}
}

\usepackage{moreverb}

\usepackage[nounderscore]{syntax}

\usepackage[pdftex]{graphicx}
\graphicspath{{./figures/}}
\DeclareGraphicsExtensions{.pdf}

\usepackage[cmex10]{amsmath}
\usepackage{amssymb}
\usepackage{mathtools}
\usepackage{amsthm}
\usepackage{amsfonts}
\usepackage{gensymb}

\usepackage{subfig} %

\usepackage{algorithmicx}
\usepackage{algpseudocode}
\usepackage[ruled]{algorithm}
\definecolor{light-gray}{gray}{0.75}
\algrenewcommand{\algorithmiccomment}[1]{\hskip3em{{\footnotesize \textcolor{light-gray}{$\blacktriangleright$}}} #1}

\usepackage{multirow} %
\usepackage{rotating} %
\usepackage{booktabs} %
\usepackage{colortbl} %
\usepackage{tablefootnote} %

\usepackage{array}
\newcolumntype{L}[1]{>{\raggedright\let\newline\\\arraybackslash\hspace{0pt}}m{#1}}
\newcolumntype{C}[1]{>{\centering\let\newline\\\arraybackslash\hspace{0pt}}m{#1}}
\newcolumntype{R}[1]{>{\raggedleft\let\newline\\\arraybackslash\hspace{0pt}}m{#1}}

\usepackage[pdftex,colorlinks=true,urlcolor=blue,citecolor=blue]{hyperref}

\usepackage{xspace}

\usepackage{enumitem}

\usepackage{blindtext}

\newcommand{\opes}{OptimES\xspace}

\def\orcid#1{\kern .08em\href{https://orcid.org/#1}{\includegraphics[keepaspectratio,width=0.7em]{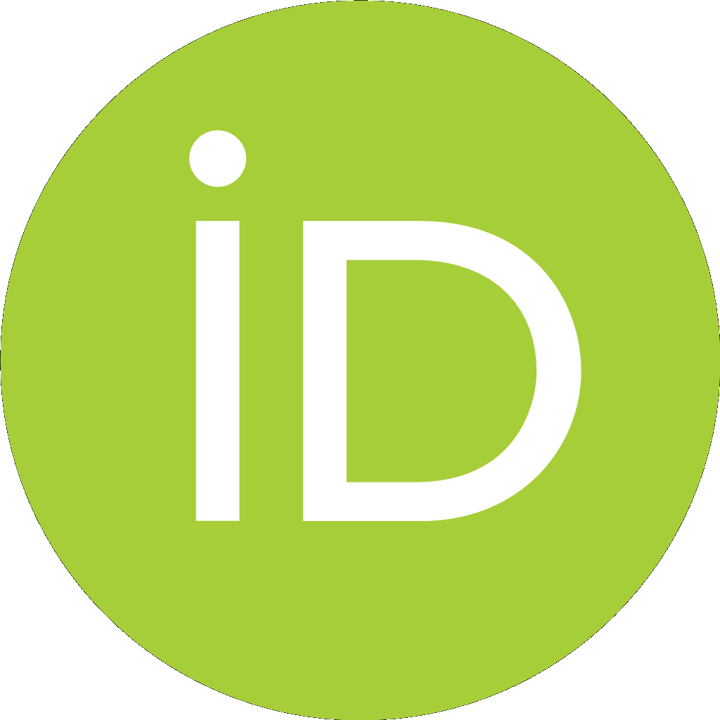}}}

\begin{document}

\title{\opes: Optimizing Federated Learning Using Remote Embeddings for Graph Neural Networks~\thanks{~Extended full-length version of paper that appeared at Euro-Par 2024: \textit{``Optimizing Federated Learning Using Remote Embeddings for Graph Neural Networks,'' Pranjal Naman and Yogesh Simmhan, in International European Conference on Parallel and Distributed Computing (Euro-Par), 2024}. DOI: \url{https://doi.org/10.1007/978-3-031-69766-1_32}}
}

\author{Pranjal Naman\orcid{0009-0000-9912-9522} and Yogesh Simmhan$^1$\orcid{0000-0003-4140-7774}\\~\\
\em Department of Computational and Data Sciences (CDS),\\
\em Indian Institute of Science (IISc),\\
\em Bangalore 560012 India\\~\\
\texttt{Email:\{pranjalnaman, simmhan\}@iisc.ac.in}
}

\date{}
\maketitle

\begin{abstract}
Graph Neural Networks~(GNNs) have experienced rapid advancements in recent years due to their ability to learn meaningful representations from graph data structures. However, in most real-world settings, such as financial transaction networks and healthcare networks, this data is localized to different data owners and cannot be aggregated due to privacy concerns.
Federated Learning~(FL) has emerged as a viable machine learning approach for training a shared model that iteratively aggregates local models trained on decentralized data. This addresses privacy concerns while leveraging parallelism. State-of-the-art methods enhance the privacy-respecting convergence accuracy of federated GNN training by sharing remote embeddings of boundary vertices through a server~(EmbC). However, they are limited by diminished performance due to large communication costs.
In this article, we propose \opes, an optimized federated GNN training framework that employs remote neighbourhood pruning, overlapping the push of embeddings to the server with local training, and dynamic pulling of embeddings to reduce network costs and training time. We perform a rigorous evaluation of these strategies for four common graph datasets with up to $111M$ vertices and $1.8B$ edges. We see that a modest drop in per-round accuracy due to the preemptive push of embeddings is out-stripped by the reduction in per-round training time for large and dense graphs like Reddit and Products, converging up to $\approx 3.5\times$ faster than EmbC and giving up to $\approx16\%$ better accuracy than the default federated GNN learning. While accuracy improvements over default federated GNNs are modest for sparser graphs like Arxiv and Papers, they achieve the target accuracy about $\approx11\times$ faster than EmbC.
\end{abstract}

\section{Introduction}
Graph Neural Networks (GNNs) have emerged as a powerful tool for learning representations on non-Euclidean data, effectively capturing dependencies and relational information inherent in them~\cite{kipf2016semisupervised}. GNNs leverage the graph topology as well as the node and edge features to learn low-dimensional embeddings, enabling them to perform tasks such as node classification, edge prediction and graph classification~\cite{kipf2016semisupervised, morris2019weisfeiler}.
GNNs find applicability in a wide range of real-world systems, including eCommerce~\cite{zhu2019aligraph}, social media~\cite{ying2018pinterest}, Google Maps~\cite{derrow2021google}, drug discovery~\cite{gaudelet2021drugdiscovery} and molecular property predictions~\cite{wieder2020molecular}.

When training a \textit{k}-layered GNN architecture, the \textit{forward pass} aggregates the \textit{k}-hop neighbours and their features/embeddings for each \textit{labelled training vertex} in the graph.
The \textit{backward pass} propagates the gradients with respect to the loss calculated over the predicted labels among the layers to update the weights and embeddings in the neural network.
This is done iteratively, with each \textit{mini-batch} operating on a subset of training vertices~(also called \textit{labelled/target vertices}) and their neighbourhood~(each called a \textit{computational graph}), and an \textit{epoch} of training ensuring coverage over all training vertices~\cite{hamilton2017graphsage}. In doing so, the low-dimensional embeddings generated for the nodes of the graph capture both the graph topology as well as features on the neighbouring vertices and/or edges. 

\paragraph*{Limitations of Centralized Training}
Real-world graph datasets can be substantially large, e.g., buyers and products in an eCommerce site~\cite{liu2021ecommerce}, interactions between user accounts in a fintech transaction graph~\cite{li2022fintech}, etc. This makes GNN training on a single~(even accelerated) server computationally costly and memory-intensive. Additionally, data privacy regulations for financial and healthcare entities, or more generally like GDPR,
can prohibit data owners from sharing their interaction graphs. In some organizations, graph data from business units in different regions may be present in the same cloud data center but managed independently~\cite{wang2022DPP, rajendran2021cloud}.
However, applications may wish to train GNN models over graphs that span such distributed silos. For instance, a global eCommerce seller may wish to use product purchase trends by users in one country to recommend the product to users in other countries~\cite{zhu2019aligraph}, or banks hosting their transaction graph on a fintech cloud may wish to build a common fraud model, but without revealing their graphs to each other or any central entity~\cite{zheng2021metafederated}.

\paragraph*{Federated Learning using GNNs}
Federated Learning~(FL) has recently shown promise for training Deep Neural Network~(DNN) models across data present at multiple clients~(devices, servers) without the need to move them to a central cloud~\cite{mcmahan2017fedavg,wang2022fedscope, wu2023embc, yao2023fedgcn}. FL trains DNN models locally over data present in each client, and shares these local models with a central server, which aggregates them into a single global model for one \textit{round} of training, and sends the global model back to the clients for the next \textit{round} of training. This process is repeated until the global model converges. Thus, the global model incorporates training features from data present across all the clients while avoiding sharing the original data with the server. 

Federated \textit{graph} learning extends this to graph datasets. In this article, we examine \textit{subgraph-level} GNN training over a single graph that is partitioned into subgraphs, with each subgraph present on a different client and a central aggregation server~\cite{yao2023fedgcn, zhang2021fedsage}. 
Federated GNN training spans the distributed computing continuum, from edge devices to clouds.
This article focuses on \textit{cross-silo federated learning}~\cite{sheller2020federated}, where the clients are fewer ($10$--$100$s) but more powerful and on the local area network or data center, as opposed to \textit{cross-device} federated learning, which involves many ($100$--$1000$s) less powerful edge devices like smart phones on the wide area network.
Cross-silo training is seen in transaction networks across financial or insurance institutions and enterprises that partition data based on regulatory regions. 

\begin{figure}[t!]
    \centering
    \includegraphics[trim={0cm, 4.5cm, 0cm, 2cm}, clip, width=\linewidth]{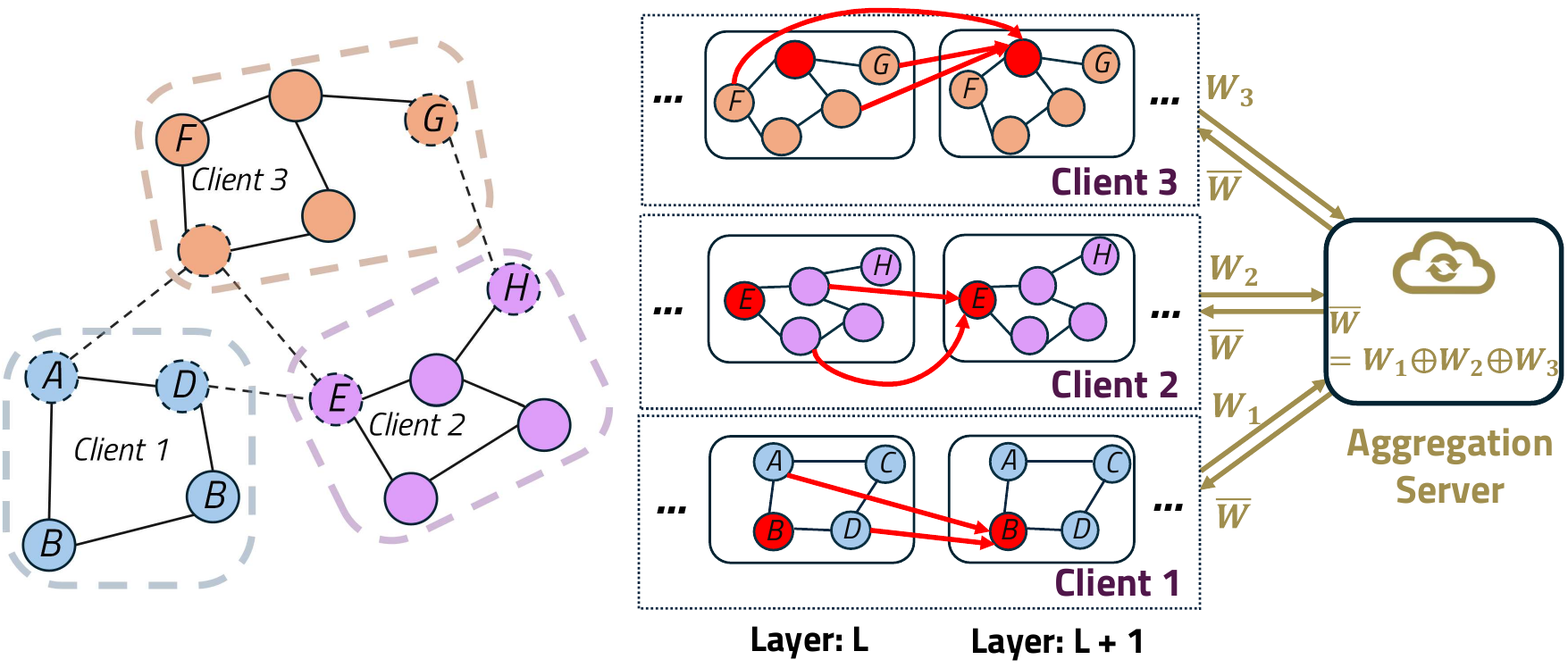}
    \caption{Design of Default Federated GNN training.}
    \label{fig:arch}
\end{figure}

Fig.~\ref{fig:arch} illustrates the \textit{default} approach to federated training of a GNN model. A local model is trained on the subgraph present in a client; these local models are aggregated on a central server, and this process repeats across rounds. 

\paragraph*{Challenges and Gaps of Federated GNN Training}
Federated learning of a GNN model over linked data presents unique challenges compared to traditional FL over independent training samples. A key limitation is the \textit{cross-client edge dependency} between subgraphs present across clients~(dashed edges in Fig.~\ref{fig:arch}),
e.g., a transaction edge between two user account vertices in two different banks. Since GNN models train over the neighbourhood information for a training vertex~(Eqn.~\ref{eq:gnn}), these cross-client edges and their downstream vertices form part of the training neighbourhood. Omitting them may result in a significant degradation of the model's performance, but including their \textit{raw features} can compromise privacy. E.g., in our experiments (Fig.~\ref{subfig:grid-1}), the Reddit graph partitioned onto $4$ clients achieves only $65\%$ training accuracy when cross-client edges are omitted but achieves $81\%$ accuracy with partial \textit{embeddings} shared, as we discuss later.

Prior works have attempted to address this limitation by sharing a subset of cross-client neighbourhood information. \textit{FedSage+}~\cite{zhang2021fedsage} proposes a generative model to generate cross-client neighbours and their features, trained simultaneously with the GNN model but at the cost of increased training time. \textit{FedGCN}~\cite{yao2023fedgcn} proposes a pre-training round to get the aggregated features of all the cross-client neighbours, but this can leak private data since the aggregated features for $1$-hop and $2$-hop are communicated directly. \textit{EmbC-FGNN}~\cite{wu2023embc} introduces a promising approach by sharing just the embeddings for the vertices incident on the cross-client edges rather than their raw features. This restricts data sharing to just the anonymized embeddings for remote vertices while allowing neighbourhood training to benefit from them. Their clients push the updated cross-client embeddings to an \textit{embedding server} after each round of training and pull the relevant ones at the start of the next round (Fig.~\ref{fig:fgnn-trainl}). This performs better than other federated graph learning strategies that they compare against. 

\begin{figure}[t]
    \subfloat[]{\label{subfig-2:inside-outside}
      \includegraphics[width=0.33\columnwidth]{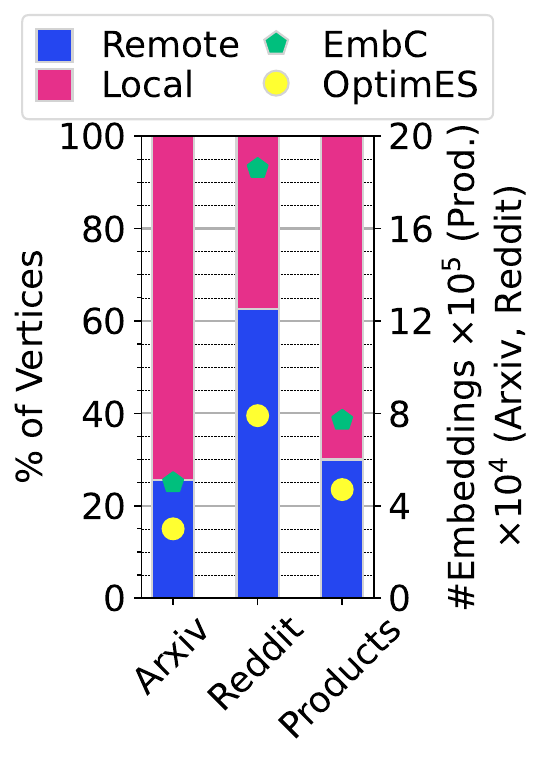}}\hfill
    \subfloat[]{\label{subfig-2:motivation3}
      \includegraphics[width=0.4\columnwidth]{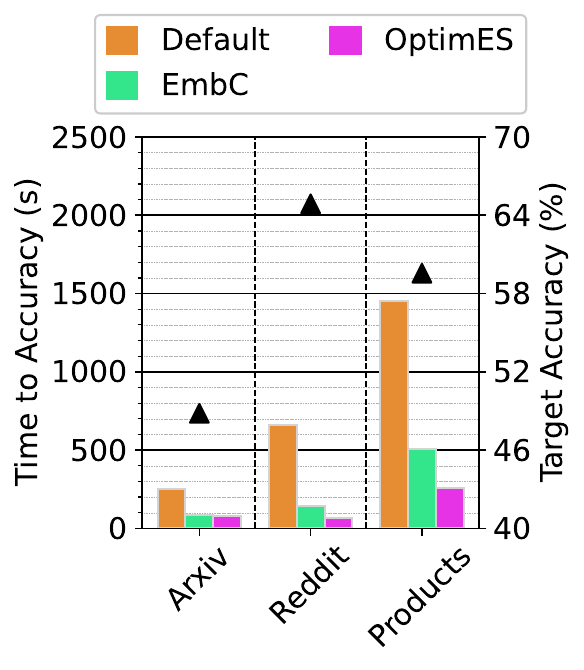}}
    \caption{(a) \% of remote vertices in subgraph partitions (left, bar) and \# of embeddings stored (right, marker). (b) Time-to-accuracy~(TTA) for our \opes over EmbC and default federated GNN training to reach the nominal peak accuracy achieved by \textbf{D}.}
    \label{fig:motivation}
\end{figure}

However, a notable drawback of EmbC is the communication cost for sharing $100,000$s of embeddings, each $200$--$400$~bytes in size, in every training round between the clients and the embedding server; this also results in a sizeable memory footprint of the embeddings held at the embedding server. 
E.g., Fig.~\ref{subfig-2:inside-outside} (bar) shows the fraction of \textit{local} and \textit{remote vertices}, i.e., unique vertices on different clients connected through a cross-client edge for three common GNN graphs we evaluate, \textit{Arxiv}, \textit{Reddit} and \textit{Products}~(each split into $4$ partitions), having $0.17$--$2.4M$ vertices and $1.16$--$123.7M$ edges. $15$--$30\%$ of their vertices are connected to a remote vertex.
For \textit{Products}, this translates to $\approx800K$ nodes for which the embedding vectors are maintained in the embedding server (Fig.~\ref{subfig-2:inside-outside}, green marker), and transferred between the server and clients in each training round. For \textit{Papers}, which is even larger and partitioned into $8$ parts, $\approx 40\%$ of vertices have cross-client edges with $40M$ embedding vectors of $\approx31$~GB in size transferred in each round.

In particular, our contributions target systems-level advancements for federated GNN training by optimizing embedding sharing, which is a critical bottleneck in distributed settings~\cite{wu2023embc}. We specifically do not claim any novel FL methods, which are better left to ML research. OptimES offers several \textit{data-level} as well as \textit{framework-level} optimizations over EmbC, allowing end users to better tailor their workload needs and balance trade-offs between performance and training speed. Our work systematically brings together the elements of graph pruning, asynchronous communication, and embedding sharing, several of which have been leveraged in isolation~\cite{liu2023comprehensive, vatter2024size, chen2023demystifying}, within an integrated framework designed for federated GNN workloads. These are the result of extensive experiments since coupling these methods \textit{ad hoc} does not accrue cumulative benefits, e.g. pulling embeddings for all~(or a subset of) remote nodes only at the start of each round is both costly and wasteful, whereas identifying and pulling only the most critical nodes initially leads to significant reductions in round time. Similarly, we also explored and compared other variants, such as static versus dynamic graph pruning~(to determine whether selecting remote nodes afresh in every round improves performance), and different staleness configurations in
overlapping communication~(to balance timeliness and bandwidth efficiency).
This integration, along with our detailed system-level evaluation, meaningfully distinguishes our work from prior systems research into federated GNN efforts and contributes new insights into the design of efficient federated GNN systems.  Lastly, these optimizations are designed with the practical consideration that not all federated workloads are alike, and therefore benefit from tunable system-level strategies.

\paragraph*{Approach and Contributions}
In this article, we propose the \textbf{Optimized Embedding Server (\opes)}, an enhanced approach that reduces the communication, computation, and memory costs for the shared embeddings method. Our strategies lead to faster training time per round, faster training time to convergence, minimum to no reduction in accuracy, and no sharing of any additional data that may compromise privacy.

Specifically, we design the following strategies: 
\begin{enumerate}[leftmargin=*]
    \item We \textit{prune} the neighbourhood of \textit{remote} vertices that are used during training through uniform sampling, reducing the embeddings transferred, the in-memory footprint, and also the compute costs for the forward and backward passes (\S~\ref{subsec:prune:rnd}). We further reduce the training time by assigning scores to the remote vertices and limiting the expansion of the local subgraph to only the high-scoring nodes (\S~\ref{subsec:prune:score}).
    \item We hide the per-round communication cost by \textit{overlapping} the transfer of embeddings from clients to the embedding server with the final epoch of training in a round, thus hiding the network costs while using slightly stale embeddings (\S~\ref{subsec:push}).
    \item We further optimize the network communication time by limiting the clients to pull only the embeddings that they will use instead of embeddings for all remote nodes, at the start of each round. We propose fetching a set of high-scoring nodes at the beginning of each round, as these nodes are certain to be useful during training. The remaining nodes are retrieved on-demand as needed.
    This reduces the overall training time, as clients only pull the embeddings for the specific remote nodes needed during training, rather than for all possible remote nodes at the start of the round~(\S~\ref{subsec:pull}). 
\end{enumerate}

We validate our strategies through detailed experiments performed on four real-world graphs, for two GNN models, GraphConv~\cite{kipf2016semisupervised} and SAGEConv~\cite{hamilton2017graphsage}, and compare them with the default and the state-of-the-art (SOTA) EmbC approaches (\S~\ref{sec:results}). 
Using \opes leads to $38\%$ fewer embeddings maintained for the Products graph~(Fig.~\ref{subfig-2:inside-outside}), 
while the time to accuracy reduces from $1460s$ for Default and $510s$ for EmbC to just $140s$ for \opes~(Fig.~\ref{subfig-2:motivation3}). We see similar significant benefits for other graphs and GNN models too.

In a prior conference paper, \textit{OpES}~\cite{naman2024optimizing}, we introduce some of these optimization strategies, viz., a random pruning strategy to limit the expansion of local subgraphs and the overlap strategy of pushing the embeddings alongside the final epoch of training.
This article, \textit{\opes}, substantially extends from the previous work to include two additional strategies: one to enhance pruning using a \textit{score-based approach} (\S~\ref{subsec:prune:score}), and another to optimize the pull phase through
\textit{scored prefetching} (\S~\ref{subsec:pull}). These further reduce the time to accuracy by up to $\approx5\times$ over OpES. Additionally, we expand our evaluation to include another GNN model~(SAGEConv) and a larger-scale graph dataset~(\textit{Papers}, with $111$M vertices and $1.6$B edges). We also conduct a scaling study as we increase the number of clients and an ablation study to find configurations when the proposed pull optimizations deliver the best performance.

In the rest of the article, we discuss related works in \S~\ref{sec:related}, present the \opes system design in \S~\ref{sec:methods},  propose our optimization strategies in \S~\ref{sec:optiz}, offer detailed comparative experiments using different configurations in \S~\ref{sec:results}, and summarize our conclusions in \S~\ref{sec:conclude}.

\section{Background and Related Work}
\label{sec:related}

\subsection{Graph Neural Networks~(GNNs)}

The ability for GNNs to learn low-level representations for graph topology and its features has found numerous applications in the real world. E-commerce platforms like Alibaba utilize GNNs to analyze user behaviour, perform personalized product recommendations and improve overall user engagement~\cite{zhu2019aligraph}. Social media platforms like Pinterest employ GNNs for content recommendation to effectively connect users with relevant content~\cite{ying2018pinterest}. Likewise, Google utilizes GNNs for the estimated time of arrival predictions in Google Maps~\cite{derrow2021google}. Complementary to these, GNNs are also used in the pharmaceutical industry for drug discovery~\cite{gaudelet2021drugdiscovery} and molecular property predictions~\cite{wieder2020molecular}.
In addition to these, they find use in a wide range of tasks such as node classification~\cite{kipf2016semisupervised} and link prediction~\cite{kipf2016variational} as well as more specialized areas like recommendation systems~\cite{ying2018graph}, and natural language processing tasks like machine translation~\cite{bastings2017graph} and relation classification~\cite{li2019classifying}.

These applications are powered by the core message passing mechanism of GNNs, where during a forward pass, a \textit{k}-layered GNN architecture aggregates the \textit{k}-hop neighbours and their features/embeddings for each training vertex in the graph.
The embedding for each layer \textit{l} for a training vertex \textit{u} is calculated using a combination of \textbf{AGGREGATE} and \textbf{COMBINE} functions as shown in Eqn.~\ref{eq:gnn}. Specifically, the previous layer $l-1$ embeddings of the neighbours of \textit{u}~($h^{l-1}_v, v \in \mathcal{N}(u)$) are aggregated, followed by a combination with the previous layer embeddings of the training vertex \textit{u}~($h^{l-1}_u$) to generate the current layer's embeddings~($h^l_u$).
In contrast, the backward pass propagates the gradients from the loss computed over the predicted labels through the layers, updating the weights and embeddings in the neural network. This process is repeated for each \textit{mini-batch} to cover all training vertices in every epoch.

\begin{equation*}
\label{eq:gnn}
    x^l_u = \textbf{AGGREGATE}^l(\{h^{l-1}_v, v \in \mathcal{N}(u)\})
\end{equation*}
\begin{equation*}
    h^l_u = \textbf{COMBINE}^l(h^{l-1}_u, x^l_u)
\end{equation*}

\subsection{Distributed GNN Training}

Various large-scale distributed training frameworks like DGL~\cite{wang2019dgl}, NeuGraph~\cite{ma2019neugraph}, BGL~\cite{liu2023bgl} and others have been proposed to address the scalability of GNN training, particularly since the entire graph and its features do not fit in GPU memory while training on CPUs is too slow. In a typical distributed GNN setting, the graph structure and/or the associated features are partitioned, with or without replication, across multiple workers and a GNN model is trained collaboratively. Specifically, the clients train the GNN model using all the data local to them, as well as by accessing data present on neighbouring clients, during an epoch. After one epoch of training~(after a mini-batch for mini-batch training), the model weights on each client are aggregated using a parameter server or a distributed all-reduce pattern~\cite{li2020ddp}. While closely related, federated GNN training differs from traditional distributed GNN training in that clients have limited or no access to the data held by other clients. Also, training happens over multiple federated rounds with multiple local epochs in FL, rather than multiple epochs of distributed training.

\subsection{Federated GNN Training}

Federated Learning (FL)  on graphs has gained traction recently~\cite{yao2023fedgcn, zhang2021fedsage}. This entails training a global GNN model in a decentralized manner while restricting the graph data to each data owner~(client). While the concept is quite similar to traditional FL, operating over graphs introduces an added layer of complexity. In tasks such as semi-supervised node classification, the data local to the clients may exhibit inter-client dependencies, which must be addressed in a privacy-respecting manner to achieve improved model convergence. The exchange of information to account for these inter-client dependencies can lead to large communication overheads and, if not done carefully, risk privacy leakage. 

FL on GNNs can be classified based on the graph datasets maintained on the clients: (1) \textit{Ego-network level}, where each client holds the \textit{n}-hop neighbourhood for one or more labelled vertices, e.g., interactions between the user and other entities on a smartphone client~\cite{zhang2023fedego}; (2) \textit{Subgraph level}, where each client holds a partitioned subgraph of a larger graph, e.g., transactions pertinent to one bank~(client) for a transaction graph that can span all national banks~\cite{yao2023fedgcn, zhang2021fedsage}; and (3) \textit{Graph level}, where each client stores one or more complete, independent graphs, e.g., protein molecule graphs held by multiple pharmaceutical companies~\cite{he2021fedgraphnn}.
In this article, we focus on subgraph-level GNN training.

Many methods target subgraph-based federated graph learning, particularly addressing the challenge of missing details on cross-client edges on remote clients that affect training accuracy. 
One common approach is to expand the local subgraph either by augmenting the neighbourhood or by sharing structural information among clients. The latter is common in scenarios such as social networks, where users from different continents may be connected, allowing the network topology to be shared without moving personal data~\cite{kotsios2019analysis}, and in financial networks, where institutions collaborate to detect fraud by incorporating masked neighbours from different banks into the topology without exposing private information~\cite{zheng2021metafederated}. \textit{FedSage+}~\cite{zhang2021fedsage} performs expansion of the local subgraph by generating the neighbourhood and its features through a collaboratively learned generative model. First, all clients train a neighbourhood generator model jointly by sharing gradients. The clients then use this to expand their subgraphs and perform a localized federated graph learning round. In contrast, \textit{FedGNN}~\cite{wu2021fedgnn} expands its subgraphs by augmenting the relevant nodes from other subgraphs with homomorphic encryption~(HE) to protect users’ raw features. FedGCN~\cite{yao2023fedgcn} trains Graph Convolutional Networks (GCNs) for semi-supervised node classification, allowing multiple clients to collaborate without sharing their local data every round. It significantly reduces communication costs and accelerates convergence by sharing aggregated features of up to $2$-hop remote nodes with the server in only a pre-training round. In contrast, our method, \opes, adopts an embedding sharing approach, where masked embeddings of remote nodes are exchanged each round to include cross-client neighbours in training, all while preserving data privacy.

Other works explore the challenges of data heterogeneity and propose improved aggregation techniques for topology-aware model aggregation at the server. \textit{FedPUB}~\cite{baek2023fedpub} proposes a personalized subgraph federated learning model to tackle distribution heterogeneity of data, particularly the fact that the subgraphs on data owners might belong to different communities in the global graph and a naive FL algorithm would lead to poor performance. They calculate the similarity between subgraphs and group similar subgraphs for improved performance. \textit{GCFL}~\cite{xie2021gcfl} is a federated graph classification framework that also handles the non-IID nature of the participating subgraphs and the node features. GCFL observes that most real-world graphs share at least a few similar properties, such as largest component size and average shortest path lengths, and calculates similarities among subgraphs to cluster similar clients for a better global model convergence. While \opes does not currently study the effect of data heterogeneity with respect to federated graph learning, we plan to address this as future work.

Wu et al.~\cite{wu2023embc} introduces a SOTA subgraph FL, \textit{EmbC}, based on subgraph expansion to include the latest embeddings for the cross-client neighbours at the boundary between rounds. A trusted entity called the \textit{embedding server} informs the clients of the presence of cross-client neighbouring vertices. The clients expand their local subgraphs to include these remote vertices and use them while training. Specifically,
 at the start of a round, each client \textit{pulls} the embeddings from the embedding server
 for such remote vertices that have \textit{edges to} their local vertices~(Client 2 in Fig.~\ref{fig:fgnn-trainl} pulls embeddings for nodes $\{ A, F, G \}$). Towards the end of the round, the clients also \textit{push} the updated embeddings to the embedding server for the local vertices which have \textit{edges to} vertices in other clients~(Client 2 pushes the updated embeddings for nodes $\{ E, H \}$) -- these updated embeddings will be pulled and used by those other clients in the next round~(Client 3 uses the updated embeddings of $\{ E, H \}$ in the next round). 

However, EmbC suffers overheads due to numerous embeddings being maintained in the server; this grows as the GNN model depth increases and/or the number of partitions increases, and hence, the cross-client edge cuts increase. It causes the memory footprint of the embedding server to grow~(since it holds these in-memory for fast response), and also the communication cost between client and server to grow in each round. 
Our prior work, OpES~\cite{naman2024optimizing}, partly addresses this by selecting a random pruned subset of the local subgraph and overlapping the embeddings communication and model computation within a round. This achieves faster model convergence without tangibly reducing the peak accuracy. In this article, we further extend OpES to OptimES by additional optimizations of scored graph pruning~(\S~\ref{subsec:prune:score}) and dynamic embedding fetching~(\S~\ref{subsec:pull}) which help further bring down the round time thus offering faster convergence.

\section{\opes System Design}
\label{sec:methods}

In this section, we outline our federated GNN training using an embedded server.

\subsection{Architecture}
\begin{figure}[t]
    \centering
    \includegraphics[width=0.8\columnwidth]{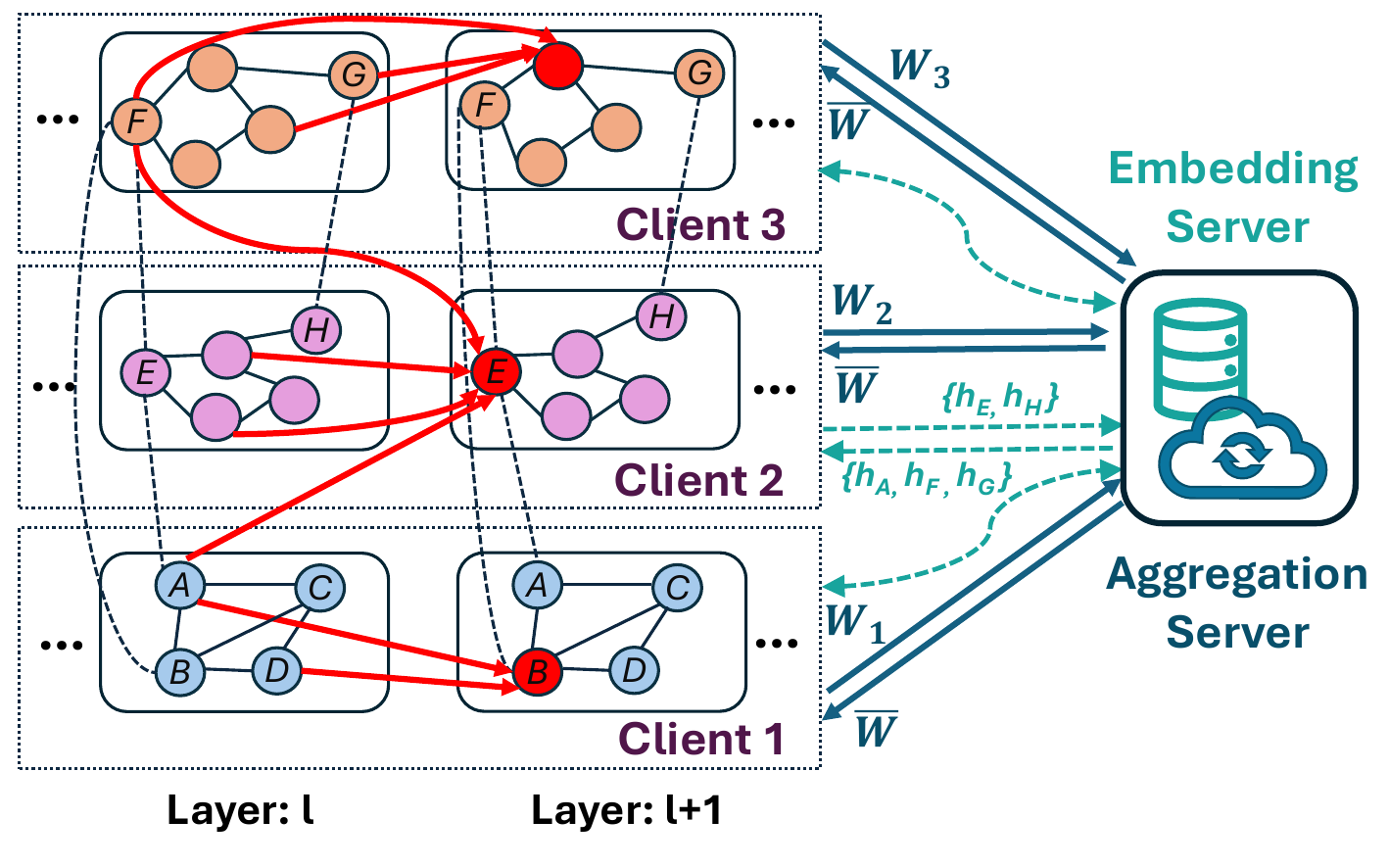}
    \caption{Federated GNN training flow using remote embeddings. Before training, each client pulls embeddings for its \textit{pull nodes}~(shown for Client 2). Training involves nodes in a mini-batch~(red nodes) aggregating neighbours' embeddings~(shown in red arrows) to generate the next layer's embeddings. After training, clients upload their local model weights to the aggregation server and the updated embeddings of their \textit{push nodes} to the embedding server. The aggregation server then aggregates these weights into the updated global model and distributes it back to all clients.}
    \label{fig:fgnn-trainl}
\end{figure}

The base architecture of \opes builds upon EmbC's base design~\cite{wu2023embc}. It consists of three key entities: a \textit{central aggregation server} that orchestrates the federated learning rounds, performs client selection, and aggregates the models; a~(possibly co-located) \textit{embedding server} that is aware of the cross-client edges and receives/sends the embeddings for the remote vertices maintained on it from/to the relevant clients; and the \textit{clients} that perform the local training rounds, interacting with the aggregation and embedding servers (Fig.~\ref{fig:fgnn-trainl}). Next, we describe these in detail.

The \textit{aggregation server} is a typical FL server that can perform client selection or model aggregation strategies such as FedAvg~\cite{mcmahan2017fedavg}, TiFL~\cite{chai2020tifl}, among others. The \textit{clients} initially possess only the local subgraph partitions of the overall graph and can query the \textit{embedding server} for the presence of cross-client neighbours~(remote vertices). Once discovered, the clients expand their local subgraphs to include these remote neighbours~(also referred to as \textit{pull nodes}) present in other clients, tagging them with a flag indicating they are remote. 

The \textit{embedding server} is an ``honest-but-curious'' entity~\cite{yang2019federated}. Only the structure of the cross-client's 1-hop vertices in the graph, i.e., just their vertex IDs, is shared with the clients; neither the embedding server nor the clients have access to the raw features of the remote vertices or the remote edges located on other clients. This standard privacy model is followed by several federated graph learning frameworks~\cite{yao2023fedgcn, wu2023embc, zhang2021fedsage}. The embedding server maintains an in-memory Key-Value~(KV) store that stores $L - 1$ embeddings~($h^1$ to $h^{L-1}$ for an $L$ layered GNN) for all vertices whose embeddings need to be shared to/from the relevant clients. The same remote vertex may be present at a $1$-hop, $2$-hop, etc. distances from a labelled training vertex that is part of a training mini-batch. So, each hop-distance translates to a different embedding for that vertex~(Fig.~\ref{fig:comp-graph:b}). As stated before, the $h^0$ embeddings~(the features) of the remote neighbours are not maintained in the embedding server.

\subsection{Lifecycle of a Training Round}
\label{subsec:lifecycle}
Fig.~\ref{fig:fgnn-trainl} shows the standard training life cycle of a federated subgraph learning using remote embeddings.
We first establish standard terminology that will be used going forward. For any client $k$, local vertices that are neighbours of another client $k'$~(i.e., remote vertices for those clients) required to train on the vertices on $k'$ are referred to as the \textit{push nodes} of client $k$. Similarly, remote vertices of client $k'$ whose embeddings are necessary to complete the training for client $k$ are termed \textit{pull nodes} of client $k$.
In general, embeddings of push nodes are expected to be ``pushed'' to the embedding server after a training round by a client, while embeddings of pull nodes are expected to be ``pulled' from the embedding server before a training round by a client, and hence the terminology.
We assume that the GNN model being collaboratively trained has $L$ layers, and we run $\epsilon$ epochs in each round of training.

\subsubsection{Pre-training}
\label{subsubsec:lifecycle:pretrain}
Before the iterative training rounds start, we run a pre-training round to initialize the embeddings for all \textit{push nodes} in each client. This occurs on the local subgraph before its expansion, calculating the $h^1$ to $h^{L-1}$ embeddings of the client's push nodes. 
This step is necessary because these embeddings are required at the very start of the first FL round, where clients begin exchanging information. This ensures that remote embeddings are already available when the FL session starts, avoiding the cold-start problem.
Client 1 in Fig.~\ref{fig:comp-graph:a} initializes the $h^1$ embeddings for $\{ A, B \}$ since these are needed by other clients. These embeddings are then sent to the embedding server for use in subsequent rounds. The pre-training round takes place once per FL session. At the end of the pre-training round, the embedding server is fully equipped with initial $L-1$ embeddings for all the remote nodes.

This initialization provides a practical and effective bootstrap, even though their calculation may initially lack global context. Subsequent training rounds progressively refine these embeddings through regular updates and embedding pulls, which incorporate cross-client structural information. This approach is common in prior systems. For example, FedGCN~\cite{yao2023fedgcn} introduces a similar pretraining phase to exchange features before training, and FedSage+~\cite{zhang2021fedsage} synthesizes neighborhood context in an offline stage using a generative model. Our system follows a similar philosophy, using this pre-training step to reduce initial communication delays and ensure a meaningful initialization of shared embeddings.

\subsubsection{Sampling and Training}
\label{subsubsec:lifecycle:sample}
\begin{figure}[t!]
\centering
      \subfloat[]{\label{fig:comp-graph:a}\includegraphics[width=0.4\columnwidth]{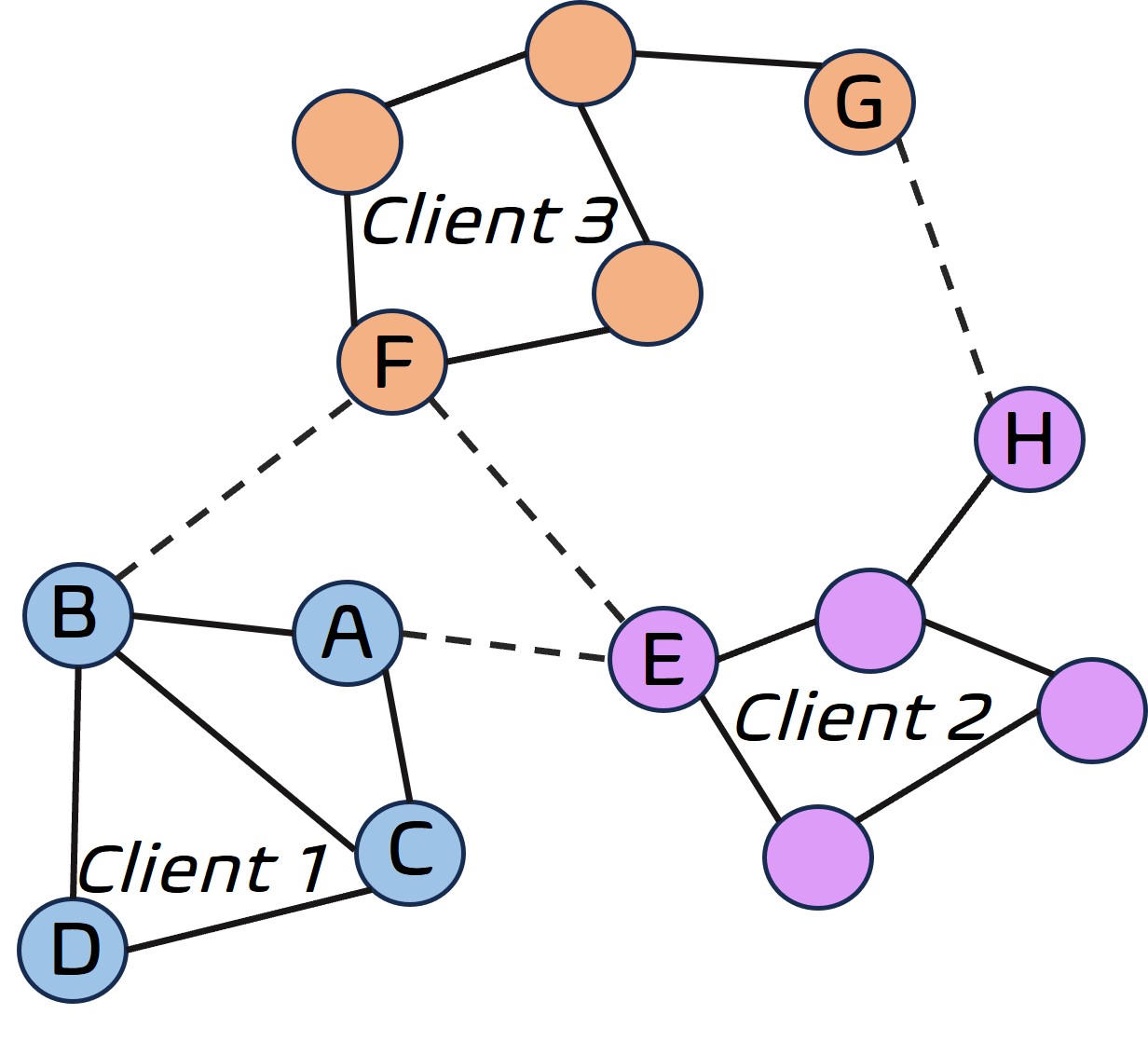}}\qquad
      \subfloat[]{\label{fig:comp-graph:b} \includegraphics[width=0.3\columnwidth]{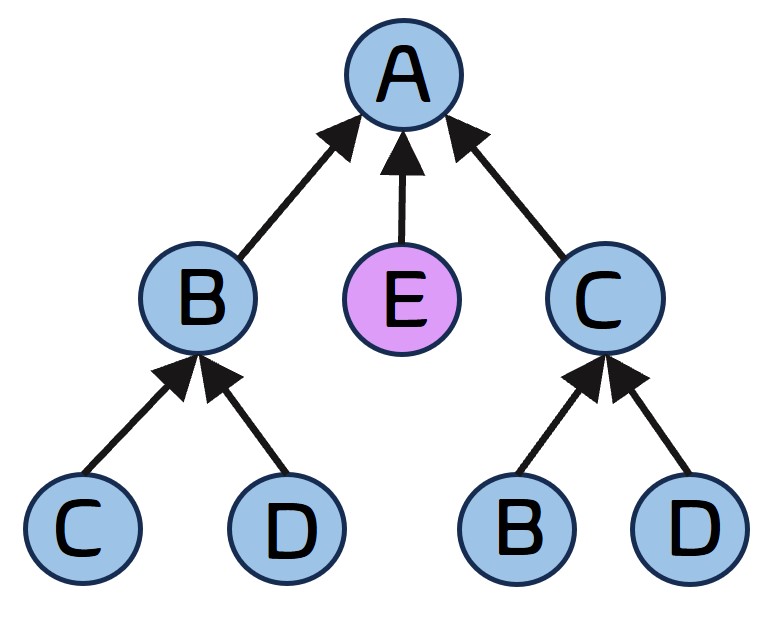}}
\caption{Existence of cross-client edges and its impact on computation graph for training node \textit{A}~(for a 2-layered GNN). (a) Subgraphs on different clients. Cross-client edges are shown as dashed edges. (b) Computation graph generated for training vertex A on client 1.}
\label{fig:comp-graph}
\end{figure}
The \textit{aggregation server} initiates a federated learning~(FL) round by selecting the participating clients for that round and sending them the global model~($\overline{W}$ in Fig.~\ref{fig:fgnn-trainl}) from the previous round~(randomly initialized model for the first round). In a \textit{cross-silo} setting, generally, all clients participate in each round. 
Each round commences with a \textit{pull phase}, where each client fetches embeddings from the embedding server for the \textit{pull nodes} to be utilized in the current round. As shown in Fig.~\ref{fig:comp-graph:a}, Client 1 pulls the $h^1 - h^{L-1}$ embeddings for nodes $F$ and $E$. Node features~($h^0$ embeddings) are unavailable at the embedding server due to privacy reasons and cannot be pulled. In this case, the GNN has $L=2$ layers; therefore, only the $h^1$ embeddings are pulled. Similarly, at the start of their individual rounds, Clients 2 and 3 pull the $h^1$ embeddings for vertices $\{ A, F, G \}$~(also shown in Fig.~\ref{fig:fgnn-trainl}) and $\{B, E, H \}$, respectively. The pulled embeddings are cached in memory locally on the client for use in this round.

GNN training is typically done in \textit{minibatches} due to the high memory requirements of full-batch training~\cite{hamilton2017graphsage}. A subset of training nodes is selected in each minibatch, and the computation graph for these is built using a \textit{neighbourhood sampler}. The sampler samples a fixed number of neighbouring nodes at each hop of the training nodes to build the computation graph.
For the \textit{sampling phase}, we develop a custom sampler that ignores the cross-client edges at the penultimate layer
and generates the computation graph such that:
(1) Only local nodes are sampled at the root level. This is the default behaviour. Between hops $1$ to $L-1$, any local or remote vertex can be sampled. However, once a remote node is sampled at hop $l \leq L-1$, the path does not grow any further, e.g., node $E$ in Fig.~\ref{fig:comp-graph:b}. (2) No remote nodes appear at the $L^{th}$ hop since the $h^0$ embeddings~(i.e., the features) of the remote nodes are unavailable. The node embeddings do not reveal any information about the specific neighbours of the remote node and only represent an aggregated view. Hence, it is safe to share embeddings among clients. 

The \textit{forward pass} of local training commences after sampling. Here, each layer \textit{l} takes the $h^{l-1}$ embeddings of the nodes at the $(L - (l - 1))^{th}$ hop as input and generates the $h^{l}$ embeddings of the nodes at $(L - l)^{th}$ hop. Before this is done, the $h^{l-l}$ embeddings for the remote vertices at the $(L - (l - 1))^{th}$ hop are loaded from the local cache. This process repeats for all the layers of the GNN to generate the $h^L$ embeddings at the output layer. The loss is calculated at the end of the forward pass and is propagated backwards along the neural network in the \textit{backward pass}. The tasks of \textit{sampling}, \textit{forward pass} and the \textit{backward pass} are repeated for all minibatches in an epoch of local training, and a total of $\epsilon$ such epochs are performed in each round.

The \textit{push phase} begins after all the epochs of a round have completed. The \textit{embedding server} must be intimated with the embeddings from the updated model for all the remote nodes. The locally trained model is used to calculate the $h^1$--$h^{L-1}$ embeddings for the \textit{push nodes} in the same manner \textit{forward pass} takes place. Note that unlike the \textit{pre-training round}, the previous round's embeddings for the \textit{pull nodes} are utilized to calculate the new embeddings of the push nodes. Once all the embeddings have been calculated for all the push nodes, these embeddings are pushed to the embedding server.

\subsubsection{Aggregation \& Validation}
\label{subsubsec:lifecycle:agg}
Immediately after the \textit{push phase}, the clients send their locally updated models to the \textit{aggregation server}, where all clients' models are aggregated to generate the global model. After aggregating updates from distributed clients, the resulting global model is evaluated on a held-out global test set at the aggregation server to measure its performance. The aggregation and validation phases take up only a minuscule portion of the total round time~(order of 100 milliseconds) since both aggregation and evaluation~(forward pass) are computationally lightweight.

\section{\opes Optimization Strategies}
\label{sec:optiz}

Here, we describe the proposed strategies to optimize the federated graph learning process using an embedded server.

\subsection{Pruning Optimizations}
\label{subsec:prune}
Full-batch training of GNNs has high memory requirements and suffers from neighbourhood explosion. There is an exponential growth in the number of vertices and edges encountered when traversing the neighbourhood of a vertex for every additional hop in a large and dense graph. To circumvent this, GNN training is generally done in mini-batches.
Hamilton et al.~\cite{hamilton2017graphsage} sample a fixed-size neighbourhood at each hop to reduce the memory requirements while still achieving great results.
Recent studies have also observed that over-smoothing significantly impacts the quality of message passing in GNNs~\cite{liu2020deepergnn}\cite{li2018deepergnn}, with the stacking of multiple GNN layers leading to notable performance degradation.
Chen et al.~\cite{chen2021dygnn} posit that this is due to redundant message passing~(some nodes converge faster and do not need deeper layers) and that not all neighbours contain task-related information~(termed \textit{edge redundancy}). E.g., when the properties of the training vertex are very similar to those of their neighbours, the neighbours do not add much value during aggregation. Additionally, including all remote nodes with only embeddings that are stale by one round can introduce more noise, leading to performance degradation.

\subsubsection{Uniform Random Pruning with Retention Limit}\label{subsec:prune:rnd}
We use the insights from the paragraph above to limit the expansion of the local subgraph to not fully encompass all remote vertices. We propose a pruning technique to \textit{restrict the maximum number of remote vertices} in the expanded subgraph. We define \textit{retention limit} as the maximum number of remote vertices a local boundary vertex can include during the subgraph expansion.
Specifically, $P_i$ is a configurable metric for vertices retained after pruning, where $i$ is the number of remote vertices retained at each vertex of the computational graph.
Doing so can accelerate the convergence time for the global model while also ensuring that the local GNN model does not completely disregard the cross-client edges. 
This reduction in time per round can arise due to multiple factors explained below.

Firstly, there is a sharp drop in the remote vertex embeddings that need to be pulled from the embedding server and cached in each training round, thereby decreasing the communication in the pull phase of the round. Additionally, the forward pass encounters fewer remote vertices and hence, has to access and populate fewer embeddings from the local cache. Lastly, pruning the expanded subgraphs also reduces the number of local push vertex embeddings that must be sent to the embedding server in a round. This pruning is performed in a uniformly random manner, i.e., a random subset of neighbouring remote vertices are removed, while staying within the retention limit.
For simplicity, our implementation does this offline before loading the subgraph.

We later demonstrate an ablation study~(\S~\ref{subsec:prune-analysis}) on pruning using different values of $P_i$.
It shows that such pruning only has a marginal effect on the peak accuracy, and yet decreases the per-round time. As is intuitive, retaining zero vertices after pruning~($P_0$) reduces this approach to a default federated GNN with no embeddings being shared; having a retention limit that is unbounded ($P_\infty$) means no vertices are pruned, and matches the baseline EmbC approach.

\subsubsection{Score-based Pruning}\label{subsec:prune:score}

To further reduce communication overhead while preserving key structural information, we propose to only include the remote nodes in the expanded subgraph that \textit{contribute the most} to the trained embeddings of labelled vertices. This is to ensure that only the most influential remote nodes are part of the expanded subgraph and we maintain the quality of the learned embeddings that have a higher relevance to the local training process.

We define a \textit{frequency score} to determine the importance of nodes.
The scoring is based on the frequency of access to a node, i.e., the number of \textit{labelled nodes} that have a particular node at an $L$-hop distance. The higher the frequency score for a \textit{pull node}, the more likely it is to appear in the computation graph of a labelled node during a forward pass and, hence, the higher its significance.

Let the \textit{L}-hop in-neighborhood of a node \textit{u} in an expanded subgraph \textit{G(V, E)} be given as:
\begin{equation*}
    \mathbb{N}_L(u) = \{ v \in V \mid d_G(u, v) \leq L \}
\end{equation*}

where $d_G(u, v)$ represents the shortest path distance along the in-edge between nodes $u$ and $v$ in \textit{G}. The \textit{frequency score} $\mathcal{S}(v)$ of a remote neighbor $v$ of node $u$ in \textit{G} with training vertex set $\mathbb{T}$ is then represented as

\begin{equation*}
    \mathcal{S}(v) = \frac{|\{x \in \mathbb{T} ~~s.t.~~ v \in \mathbb{N}_L(x)  \}|}{|\mathbb{T}|}
\end{equation*}

Before the client expands its local subgraph during \textit{pre-training} to include the \textit{pull nodes}, it computes the frequency score to each \textit{pull node} to represent its significance. 
The clients calculate the scores of the remote nodes offline using the above formula. 
The score serves as a priority list, allowing us to only include the \textit{top-f\%} of all nodes as part of the expansion.
Upon selection, the clients expand their local subgraphs during \textit{pre-training} to only include the top $f\%$ of the highest scoring nodes. For our experiments, we choose $f=25\%$.

Later, in \S~\ref{subsec:prune-analysis}, we discuss the efficacy of the scoring strategy. We perform an ablation study with increasing \textit{f} and compare the time-to-accuracy and peak accuracy metrics when retaining the top $f\%$ scoring nodes against a random $25\%$ subset of the nodes~(Fig.~\ref{fig:revision-plot-1}). We also introduce two other scoring metrics based on \textit{bridge-centrality}~\cite{jones2021bridge} and \textit{degree-centrality}~\cite{zhang2017degree} to compare against our scoring strategy. The centrality-based methods include exchange of centrality scores among clients in the \textit{pre-training} phase~(\S~\ref{subsubsec:lifecycle:pretrain}), using which the clients score and prune their remote nodes. We note here that the centrality-score-based pruning methods follow a more relaxed privacy model than our frequency scoring strategy. Additionally, the calculation of centrality scores can be localized to a client's subgraph and can be done offline, and thus does not contribute towards FL round times.

\begin{figure}[t]
    \centering
    \includegraphics[trim={0cm, 5cm, 9cm, 2cm}, clip, width=0.85\linewidth]{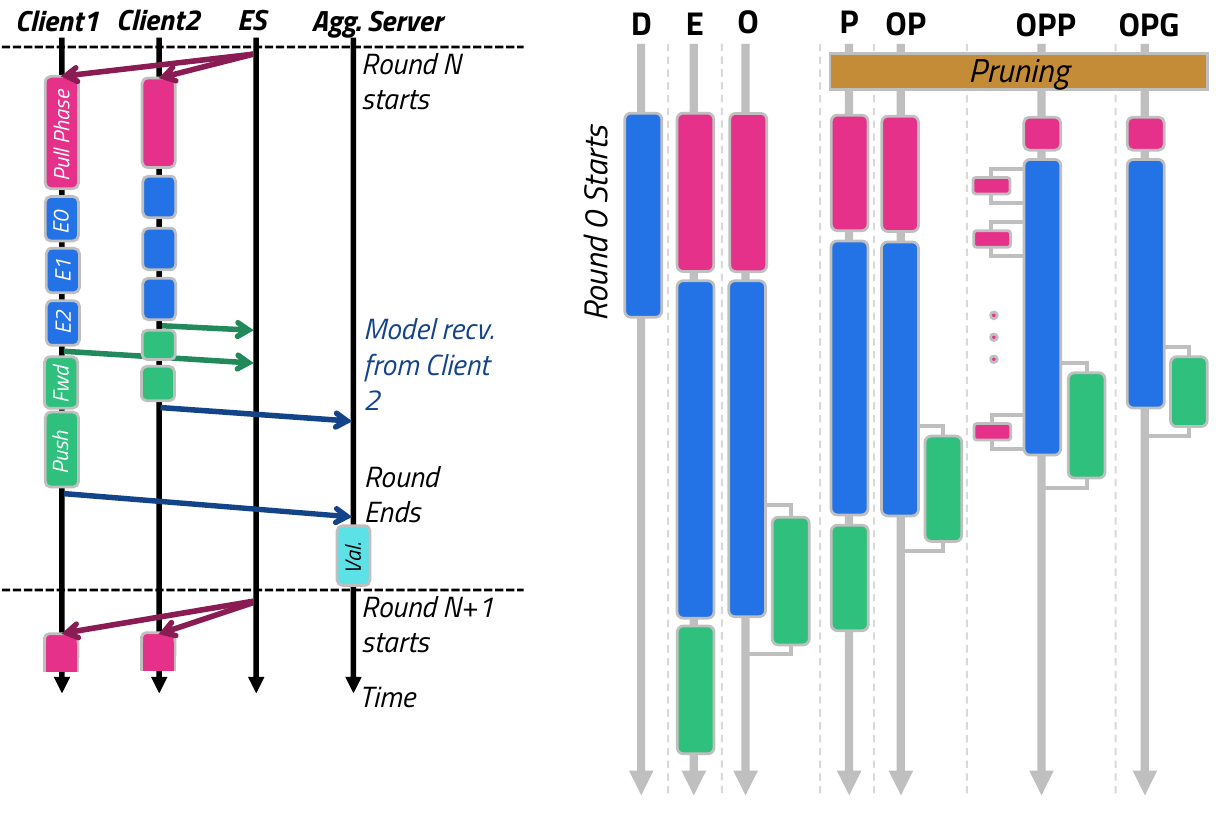}
    \caption{Generic timeline diagram for embedding-sharing federated GNN training (left) and its specialization for the proposed strategies (right). Pruning takes place offline on the clients, for simplicity.}
    \label{fig:seq-diag}
\end{figure}

\subsection{Overlapping Push Phase with Training}
\label{subsec:push}
Another key strategy we propose further hides the time taken by the \textit{push phase} of a round. A round typically consists of multiple epochs of local training so that the local model stabilizes its learning~(see left timeline in Fig.~\ref{fig:seq-diag}). We use the intuition that the embeddings for the push vertices that are computed are unlikely to change by much between the end of the penultimate epoch $(\epsilon - 1)$ and the last epoch $\epsilon$. This gives us the opportunity to
asynchronously send the embeddings for the push vertices' states at the end of epoch $(\epsilon - 1)$ while training of the final epoch is ongoing. The push is communication bound while the training is compute bound; hence, this overlapping of push and training phases allows complementary resources to be used and the push phase time to be hidden within the training phase time, albeit with a slightly stale version of the embeddings. 

Specifically, at the end of the epoch $(\epsilon - 1)$ epoch in a round, we launch a separate process on each client that takes the trained local model until that epoch and performs a forward pass to generate the embeddings for layers $h^1$--$h^{L-1}$ of the push vertices. It then sends these to the embedding server. The client will concurrently continue to train for the local model for the last epoch $\epsilon$ in the round, which will be sent to the aggregation server for global model aggregation. This optimization is relevant only when $\epsilon \geq 2$; in our experiments, $\epsilon=3$.

\subsection{Pull Phase Pre-fetching}
\label{subsec:pull}

Considering this, it becomes critical to devise strategies to optimize the time taken by this phase.

One obvious idea is to check whether all pull nodes for a subgraph fall within \textit{L}-hops of the training vertices on that client since only they will ever be used during training. But we notice that even for a small $3$-layered GNN, the $L$-hop neighbourhood of the training vertices for all four datasets in our experiments encompass $98\%$--$100\%$ of the \textit{pull nodes}. So it is not beneficial 
to avoid pushing/pulling remote vertices that fall outside the neighbourhood of the training vertices.

However, we note that only a fraction of these are actually used during training due to two reasons: 
(1) While embeddings are computed and pushed for all layers of a remote vertex ($h^1...h^{L-1}$), the vertex itself may never be at every distance $1...(L-1)$ from any training vertex, e.g., $h^{L-1}$ is used but $h^{L-2}$ is never used.
(2) Since we sample the vertex neighbourhood when constructing the computational graph for each minibatch, we use the embeddings of only a small subset of vertices within the $L$ hop neighbourhood.
We notice that in Reddit, only $40\%$ of $h^2$ embeddings are pulled and only $60\%$ of remote vertices within the neighbourhood are sampled.
So, retrieving and caching embeddings for all potential pull nodes during the pull phase is unnecessary.

One possibility is to only fetch the relevant embeddings for the pull nodes that are part of the computation graph for a training vertex during the forward pass of a minibatch. 
However, fetching all these embeddings incrementally can be costlier due to the overhead of many RPC calls.
Instead, we balance between pulling \textit{all embeddings at the start} in a single batched invocation, which has redundancy, and pulling \textit{only embeddings used in each minibatch on-demand}, which makes many small calls, by pulling \textit{nodes highly likely to be used in batch mode} at the start with a greater response throughput from the embedding server, and \textit{incrementally pulling the rest that are used, on-demand}.
Specifically, we identify the important pull nodes in the expanded subgraph based on their frequency score in \S~\ref{subsec:prune:score}.
We preemptively pull the embeddings for only a fraction $x$ of the best-scoring nodes in the pull phase.
The other pull nodes that are required during training are pulled on-demand. While this increases the time spent in the train phase, the overall reduction in pull phase time is more significant, resulting in a lower total round time.

We select this fraction $x\%$ of high-scoring nodes to pull depending on various factors such as the distribution of node scores, batch size, etc. 
The minibatch size dictates the maximum number of requests to the embedding server, with at most one request made per batch. If many vertices have a higher score, then  
pulling a small fraction initially can still result in a large fraction of the rest still being pulled during training as they are likely to be sampled.
Conversely, if a few vertices have high scores, 
then pulling these smaller fractions of nodes initially may account for most of the embeddings used. In our experiments, we set $x=25\%$ based on an ablation study. We also report results for the study where we demonstrate the effectiveness of the scoring strategy for different values of $x$.

The earlier score based pruning (\S~\ref{subsec:prune:score}) is a variant of this approach. There, we statically prune the graph to only use the $25\%$ of high-scoring vertices and nothing else subsequently, potentially impacting the accuracy. Here, we initially include the high-scoring vertices and later fetch the missing embeddings on-demand; this improves the performance without affecting the accuracy. These are complementary strategies.

\section{Experiments and Results}
\label{sec:results}

\subsection{Implementation}
We implement the default federated GNN and the EmbC-FGNN strategy, along with all our optimizations, using DGL~\cite{wang2019dgl}, Pytorch and our Flotilla framework~\cite{banerjee2025flotilla}. DGL provides the necessary APIs to implement the local training of GNN models, while PyTorch enables other tensor operations, such as maintaining the local embeddings cache. Additionally, DGL leverages PyTorch as its backend to enable the forward and backward passes during model training.
we use our Flotilla federated learning framework for orchestration of model aggregation and local training.
Once a client is selected, the Flotilla leader invokes local training on the client using DGL, which in turn initiates the pull phase. After all necessary embeddings are pulled, the trainer completes the local model training and then initiates the push phase, followed by a push of the local model to the aggregation server. Once the Flotilla leader receives the updated local models from the clients, it aggregates them into a global model and performs its validation using the test dataset to record its accuracy.

The embedding server is implemented as a Redis server that stores $\langle key, value \rangle$ pairs, with the $key$ being the node ID and the $value$ being the corresponding embedding. We maintain a separate database for each layer's embeddings~($L-1$ in total) 
to allow scoped updates. 
Redis \textit{get} and \textit{set} Remote Procedure Calls (RPC) 
are inefficient when handling individual requests for $L-1$ embeddings each for $100,000$s of remote nodes. To avoid this we use batched requests combined with pipelining. This approach allows us to queue multiple commands and send them to the server in a single batch, significantly reducing latency for RPC requests for each remote node.


\begin{figure*}[t]
    \centering
    \subfloat[Reddit TTA and Peak accuracy\label{subfig:grid-1}]{\includegraphics[width=0.4\textwidth]{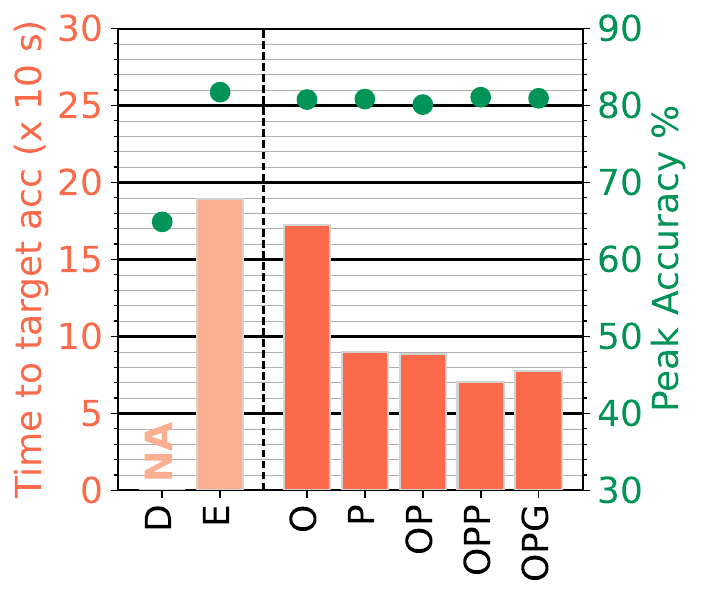}} \qquad
    \subfloat[Products TTA and Peak accuracy\label{subfig:grid-2}]{\includegraphics[width=0.4\textwidth]{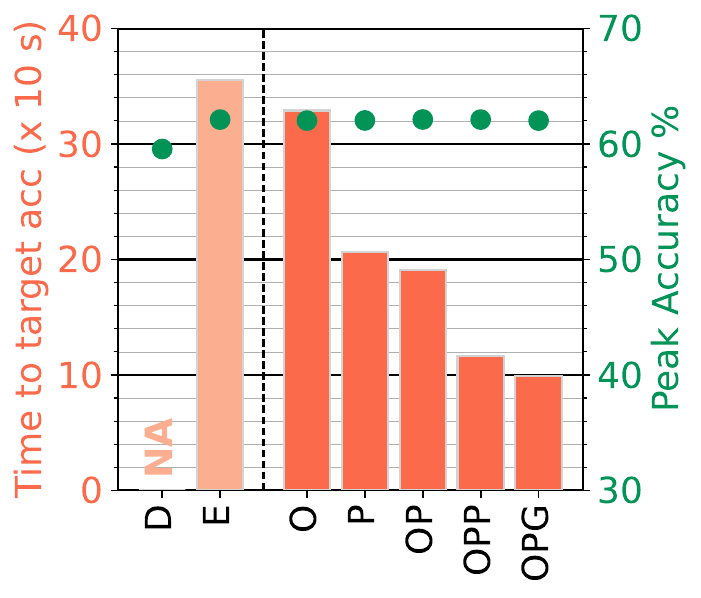}}\\
    \subfloat[Arxiv TTA and Peak accuracy\label{subfig:grid-3}]{\includegraphics[width=0.4\textwidth]{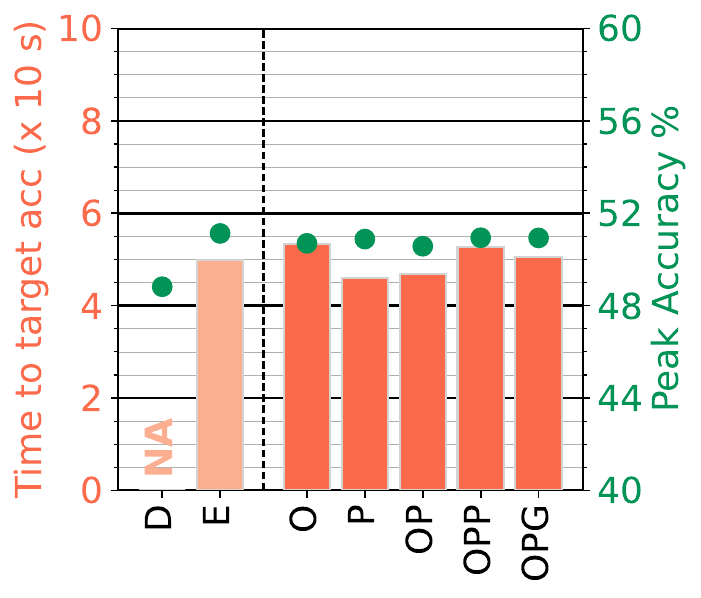}} \qquad
    \subfloat[Papers TTA and Peak accuracy\label{subfig:grid-4}]{\includegraphics[width=0.42\textwidth]{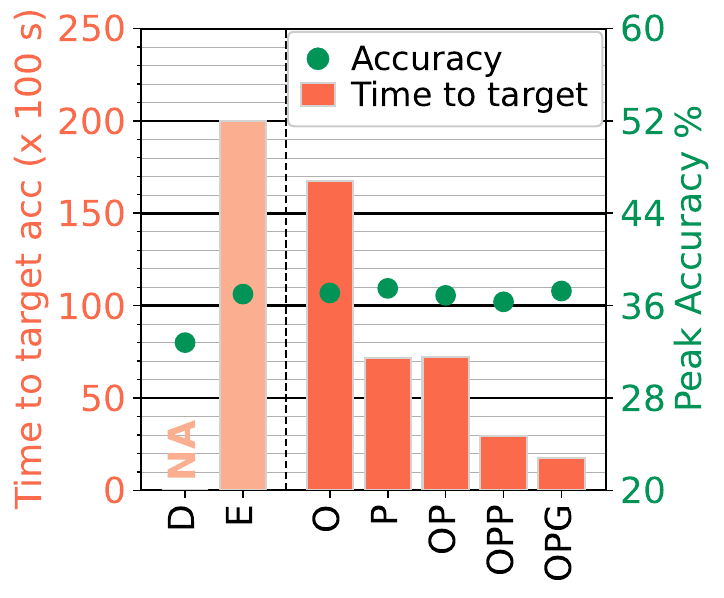}} \\
    \caption{Time to Target Accuracy (TTA)~(left y-axis) and Peak accuracy~(right y-axis) for a $3$-layered GraphConv.}
    \label{fig:grid-gc-1}
\end{figure*}

\begin{figure*}[t]
    \centering
    \subfloat[Reddit Round Time\label{subfig:grid-5}]{\includegraphics[width=0.4\textwidth]{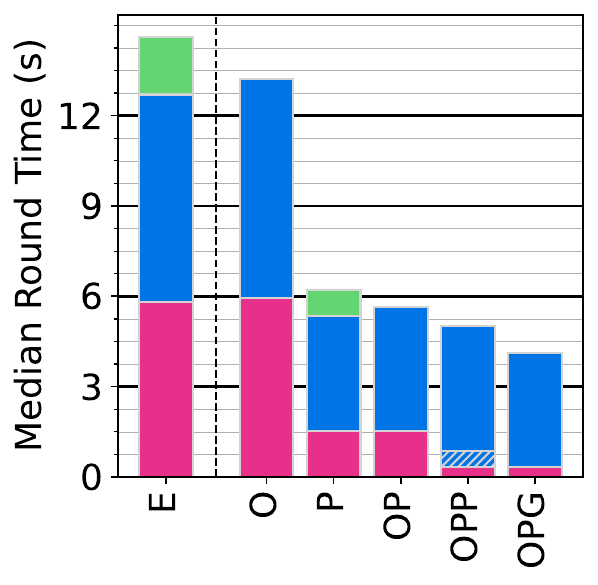}} \qquad
    \subfloat[Products Round Time\label{subfig:grid-6}]{\includegraphics[width=0.4\textwidth]{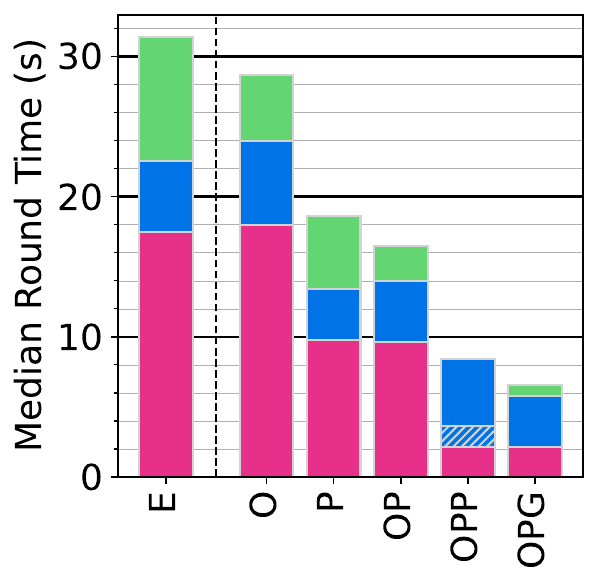}} \\
    \subfloat[Arxiv Round Time\label{subfig:grid-7}]{\includegraphics[width=0.4\textwidth]{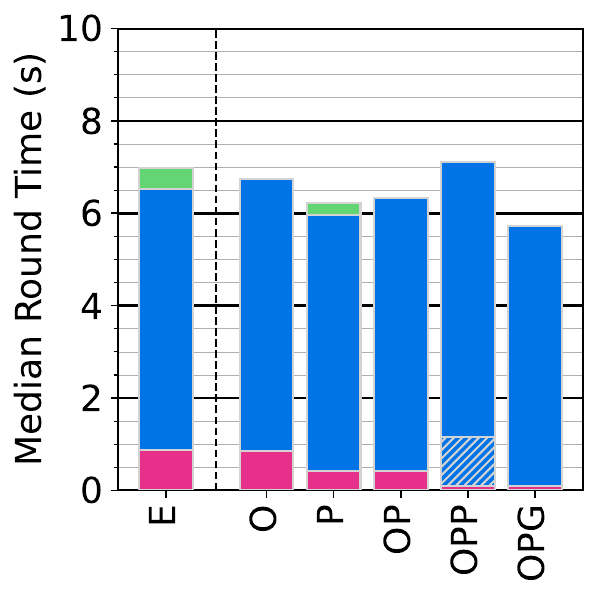}} \qquad
    \subfloat[Papers Round Time\label{subfig:grid-8}]{\includegraphics[width=0.4\textwidth]{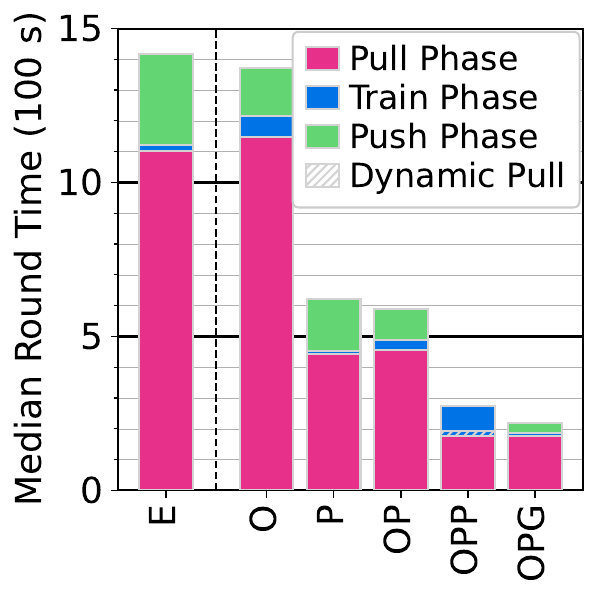}} \\
    \caption{
    Median round time and its components for pull, train, and push phases for a $3$-layered GraphConv. 
    }
    \label{fig:grid-gc-2}
\end{figure*}

\begin{figure*}[t]
    \centering
    \subfloat[Reddit Accuracy\label{subfig:grid-9}]{\includegraphics[width=0.4\textwidth]{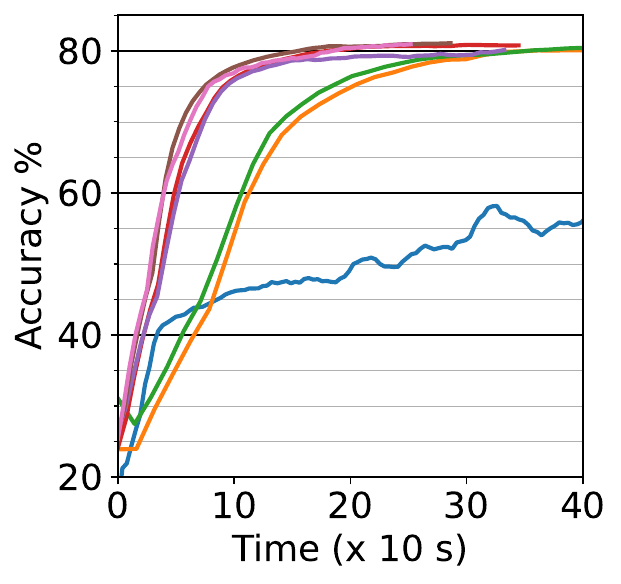}} \qquad
    \subfloat[Products Accuracy\label{subfig:grid-10}]{\includegraphics[width=0.4\textwidth]{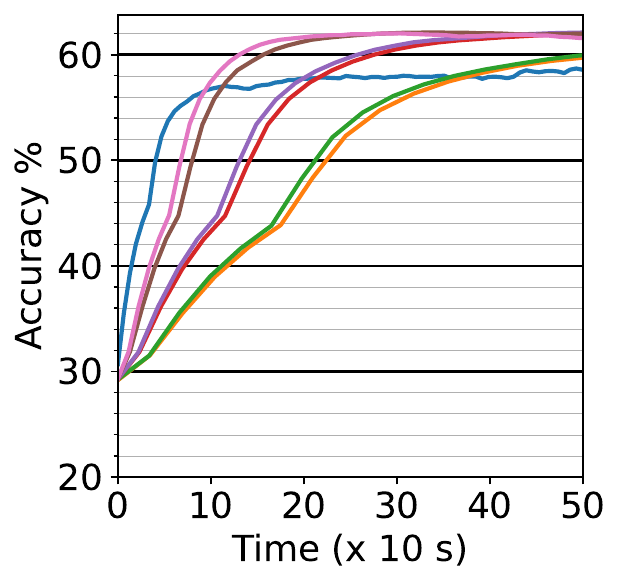}} \\
    \subfloat[Arxiv Accuracy\label{subfig:grid-11}]{\includegraphics[width=0.4\textwidth]{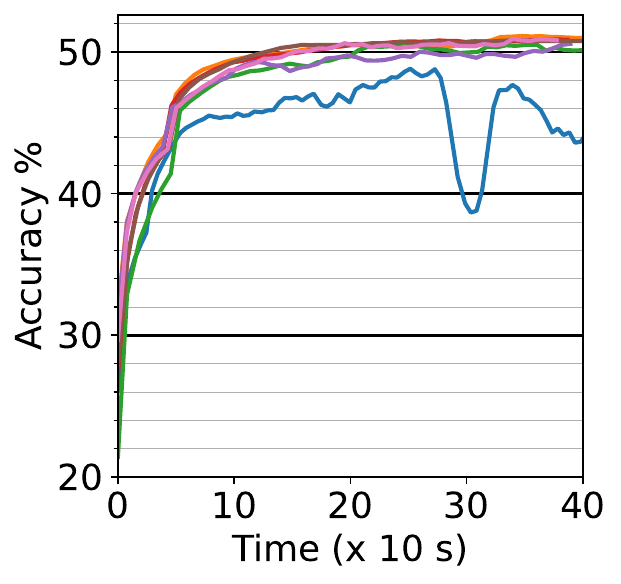}} \qquad
    \subfloat[Papers Accuracy\label{subfig:grid-12}]{\includegraphics[width=0.4\textwidth]{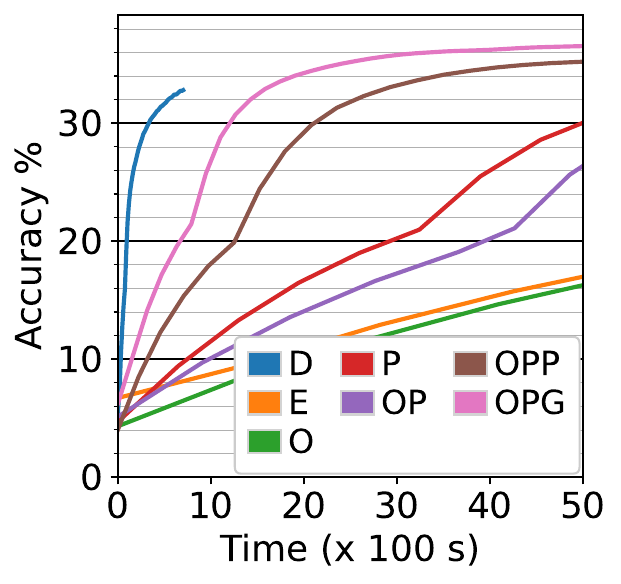}}
    \caption{
    Convergence of accuracy (avg. of $5$ rounds), for a $3$-layered GraphConv.
    }
    \label{fig:grid-gc-3}
\end{figure*}

\subsection{Experimental Setup}
\paragraph*{Hardware Setup}
Our experimental setup consists of eight GPU workstations that host the \textit{clients}, each having an Nvidia RTX 4090~(24GB) GPU card and an AMD Ryzen 9 7900X with 12-Core CPU with $128$GB RAM. 
Each client workstation runs a Flotilla client instance to manage the local training.
For simplicity, the Redis store with embeddings and the Flotilla aggregation service both run on a server with an Nvidia RTX A5000~(24GB) GPU card, an AMD EPYC 7532 $32$-Core CPU and $512$GB of RAM.
All machines are connected using a 1~Gbps Ethernet interface.

\begin{table}[t]
\centering
\footnotesize
\caption{Graph datasets used in experiments.} 
\begin{tabular}{l||r|r|r|r|r|r}
\hline
\textbf{Graph} & $|V|$  & $|E|$   & Feats. & \# Classes & Avg. In-Deg. & Train Verts. \\ \hline\hline
\textbf{\textit{Arxiv}}~\cite{hu2020open}                & 169K   & 1.2M  & 128 & 40 & 6.9 & 90.9K \\
\textbf{\textit{Reddit}}~\cite{hamilton2017graphsage}               & 233K   &  114.9M  & 602 & 41 & 492 & 153.4K \\
\textbf{\textit{Products}}~\cite{hu2020open}         & 2.5M   & 123.7M & 100  & 47 & 50.5 &  196.6K\\ 
\textbf{\textit{Papers}}~\cite{hu2020open}          & 111M   & 1.62B & 128 & 172 & 14.5 &  1.2M\\  \hline
\end{tabular}
\label{tab:dataset-specs}
\end{table}

\paragraph*{Graph Datasets}
We use four popular GNN datasets for our experiments: \textit{Arxiv} \cite{hu2020open}, \textit{Papers100M}~\cite{hu2020open} and \textit{Products}~\cite{hu2020open} from the OGBN repository, and Reddit~\cite{hamilton2017graphsage} from DGL data repository. The first two represent citation networks, while Products has data on an eCommerce website and Reddit consists of social network posts~(Table~\ref{tab:dataset-specs}). We divide the large Papers graph into $8$ subgraphs, running on $8$ client workstations, while the others are partitioned into $4$ subgraphs on $4$ clients. We use the METIS partitioner~\cite{karypis1997metis} with vertex balancing and minimized edge cuts.

\paragraph*{GNN Models}
Our primary experiments use a $3$-layer GraphConv~\cite{kipf2016semisupervised} GNN model with a hidden embedding size of $32$. For completeness, we also evaluate the proposed strategies when training a 3-layer SAGEConv~\cite{hamilton2017graphsage} model. The configurations for both models are the same. We perform neighbourhood sampling at each layer with a fanout of $5$, i.e., for each node, $5$ neighboring nodes are sampled when constructing the computation graph.
The epochs per round are set to $\epsilon=3$, and the learning rate to $0.001$. The batch sizes selected for Arxiv, Reddit, Products and Papers are $64$, $1024$, $2048$, and $4096$, respectively. 
We set $x=25\%$ for the frequency scoring strategies. All training experiments are run for $50$ rounds.

\paragraph*{Metrics and Notations}
The accuracies are measured on a global~(held out) test dataset present on the aggregation server during the aggregation and validation phase~(\S~\ref{subsubsec:lifecycle:agg}). 
We use a combination of time-to-accuracy, median round times and accuracy convergence metrics to show the efficacy of the proposed optimizations. The time-to-accuracy is the time required for the GNN model to reach a predefined target accuracy, which is within $1\%$ of the minimum peak accuracy achieved across the strategies being compared.
For the convergence plots, we report a moving average over $5$ rounds to smooth out transient spikes. Finally, for all except the retention experiments (Fig.~\ref{fig:peak-acc-prune}), the $P_4$ configuration is used for uniform random pruning.

For brevity, we use the following acronyms for the seven strategies in our plots and analysis: \textbf{D} for \textit{\textbf{\underline{D}}efault federated GNN}
and \textbf{E} for \textit{\textbf{\underline{E}}mbC}, which form the baselines. 
For the \opes variants, we use 
\textbf{O} for \textit{Push \textbf{\underline{O}}verlap Optimization};
\textbf{P} for uniform \textit{\textbf{\underline{P}}runing Optimization};
\textbf{OP} for \textit{\textbf{\underline{O}}verlap} and uniform \textit{\textbf{\underline{P}}runing} combined;
\textbf{OPP} for 
\textit{\textbf{\underline{OP}}} coupled with
\textit{\textbf{\underline{P}}refetch}; and 
\textbf{OPG} for 
\textit{\textbf{\underline{OP}}} combined with
\textit{Scored \textbf{\underline{G}}raph Pruning}.

\begin{figure*}
    \centering
    \subfloat[Reddit TTA and Peak accuracy%
    \label{subfig:gs1}]{\includegraphics[width=0.46\textwidth]{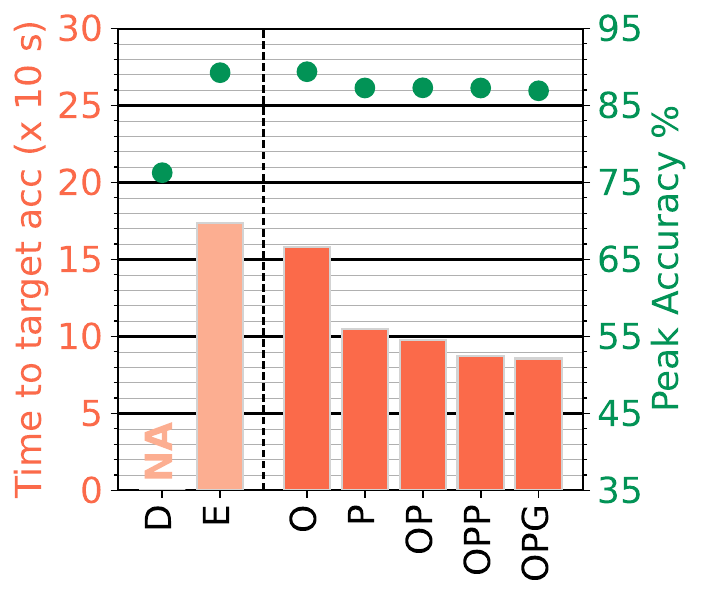}}%
    \qquad
    \subfloat[Reddit Round Time\label{subfig:gs4}]{\includegraphics[width=0.4\textwidth]{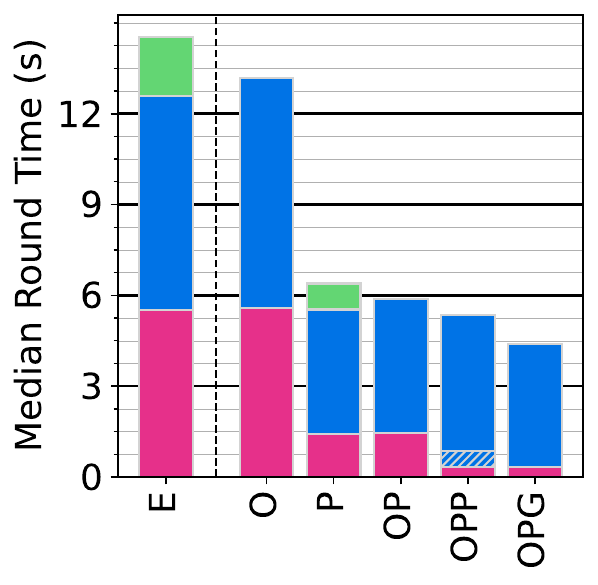}} \\
    \subfloat[Products  TTA and Peak accuracy]%
    {\label{subfig:gs2}
    \includegraphics[width=0.46\textwidth]{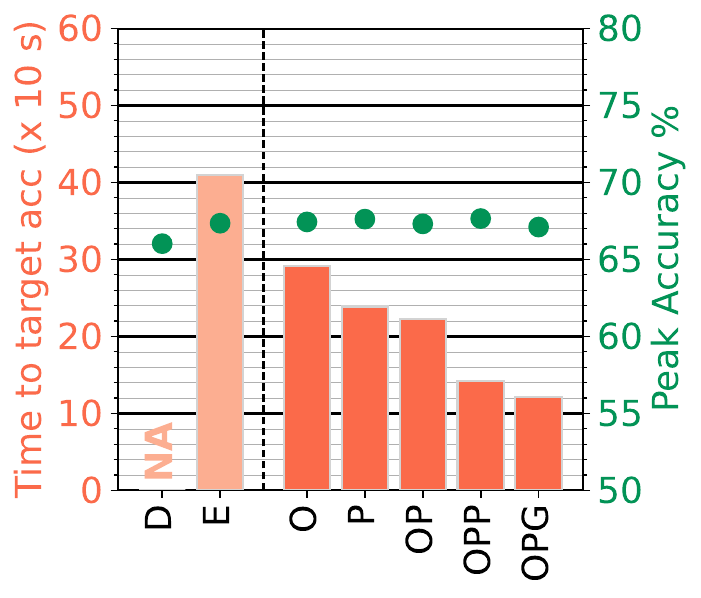}}%
    \qquad
    \subfloat[Products  Round Time\label{subfig:gs5}]{\includegraphics[width=0.4\textwidth]{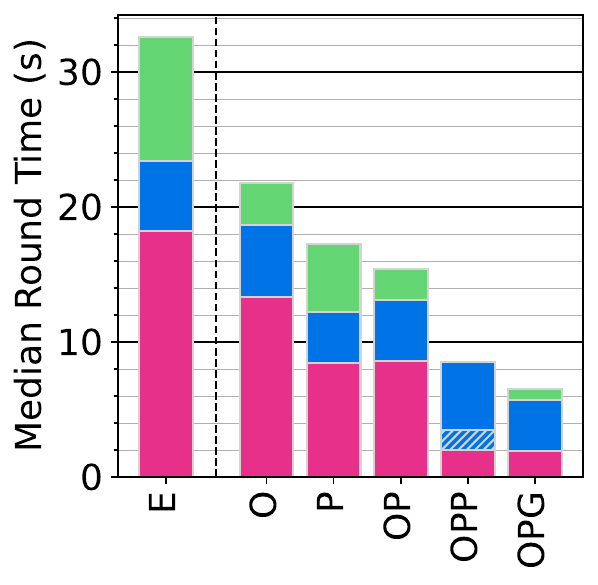}}\\
    \subfloat[Arxiv TTA and Peak accuracy]%
    {\label{subfig:gs3}\includegraphics[width=0.46\textwidth]{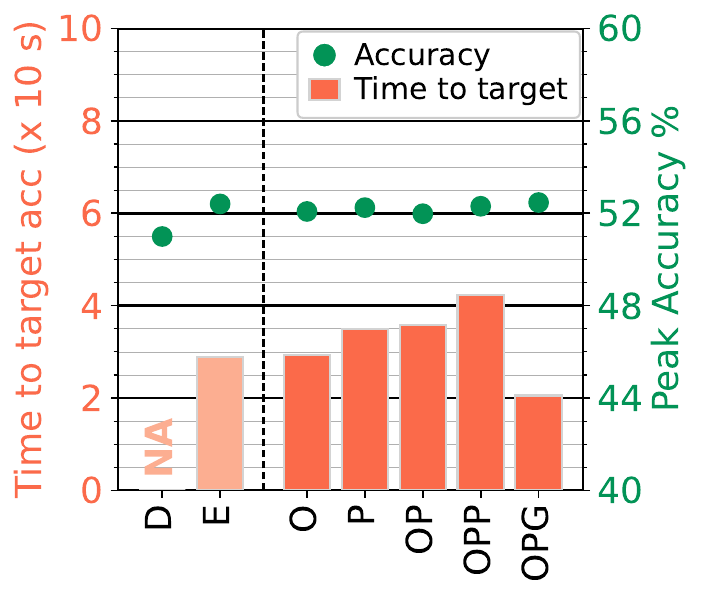}}%
    \qquad
    \subfloat[Arxiv  Round Time\label{subfig:gs6}]{\includegraphics[width=0.4\textwidth]{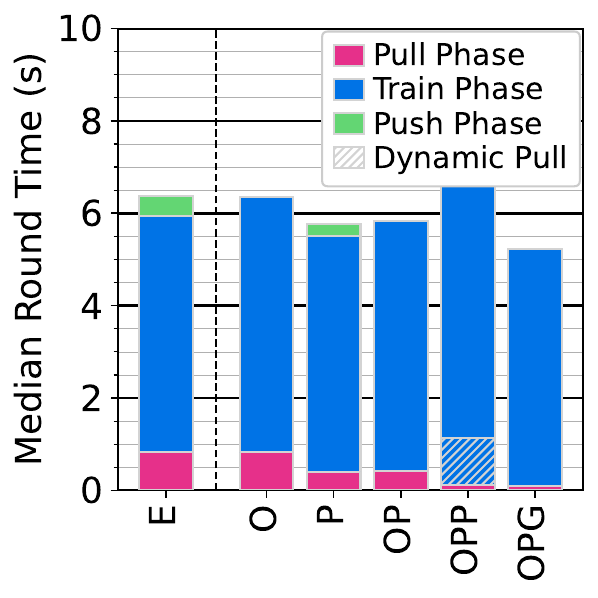}} \\
    \caption{Time to Target Accuracy (TTA)~(left y-axis) and Peak accuracy~(right y-axis) in the left column; Median round time and its components for pull, train and push phases~(right column), for a $3$-layered SAGEConv.}
    \label{fig:grid-gs}
\end{figure*}

We first offer high-level observations on the performance of the model training on the four graph datasets using the two baselines and five \opes strategies. We later examine the detailed impact of each strategy and the effects of client scaling and other that affect our optimizations.

\subsection{Comparison with Baselines}
Figures~\ref{fig:grid-gc-1}--\ref{fig:grid-gc-3} and~\ref{fig:grid-gs} summarize the results of all proposed \opes optimizations compared to the \textbf{D} and \textbf{E} baselines for the GraphConv and SAGEConv GNN model training, respectively. Fig.~\ref{fig:grid-gc-1} shows the time-to-accuracy and peak accuracy, Fig.~\ref{fig:grid-gc-2} plots the corresponding round times, and Fig.~\ref{fig:grid-gc-3} has accuracy convergence plots across rounds for GraphConv. Fig.~\ref{fig:grid-gs} shows the time-to-accuracy and peak accuracy in the left column and the corresponding round times in the right column for SAGEConv.
This is shown for all 4 graphs, Reddit, Products, Arxiv, and Papers for GraphConv, and 3 graphs except Papers for SAGEConv.

\subsubsection{Reddit}

Reddit sees a sharp increase in the peak accuracy for EmbC and for the \opes strategies over the \textbf{D} baseline, improving by $\approx16\%$ for GraphConv~(Fig.~\ref{subfig:grid-1}) and $\approx13\%$ for SAGEConv~(Fig.~\ref{subfig:gs1}).  Further, the \opes strategies are able to outperform EmbC~(\textbf{E}) for GraphConv on \textit{time-to-accuracy}~(TTA). We see a steady reduction in TTA of $\approx2.2\times$, $\approx2.5\times$, and $\approx2.4\times$ for \textbf{OP}, \textbf{OPP}, and \textbf{OPG} over EmbC~(Fig.~\ref{subfig:grid-1}), confirming the benefits of the increased optimizations.
The corresponding reduction in TTA for SAGEConv are $\approx1.8\times$, $\approx2\times$, and $\approx2.1\times$~(Fig.~\ref{subfig:gs1}) for \opes strategies over EmbC.

This is a consequence of significant drops in the median per-round time~(Fig.~\ref{subfig:grid-5}) for \opes when training for GraphConv, of $\approx2.5\times$, $\approx2.9\times$, and $\approx3.5\times$ for \textbf{OP}, \textbf{OPP}, and \textbf{OPG} strategies, respectively, compared to EmbC~(similar benefits for SAGEConv are seen in Fig.~\ref{subfig:gs4}). 
Even though the median round times reduce the most for \textbf{OPG}~(Fig.~\ref{subfig:grid-5}), it takes more training rounds to converge to the target accuracy~(Fig.~\ref{subfig:grid-1}).
This is likely due to information loss caused by pruning the Reddit graph to include only the top $25\%$ of pull nodes in \textbf{OPG}.
Since Reddit is a dense graph with many highly-scoring pull nodes, limiting the graph to just $25\%$ of these nodes results in the exclusion of many other high-scoring nodes.
These can be validated through the accuracy convergence plots in Fig.~\ref{subfig:grid-9}, where \textbf{OPP} (brown line) converges the fastest.

\subsubsection{Products}

Overall, we see a modest increase of $\approx2.6\%$ and $\approx1.6\%$ in the peak accuracy achieved by \textbf{E} and other \opes optimizations over \textbf{D}, for GraphConv and SAGEConv, due to their use of remote embeddings.

The benefits of the pull optimization \textbf{OPP} is much more pronounced here, compared to Arxiv or Reddit (Fig.~\ref{subfig:grid-2}) -- we see a sharp drop in TTA from \textbf{OP} to \textbf{OPP}~($\approx1.8\times$ as compared to Reddit's $\approx1.3\times$ and Arxiv's $\approx0.9\times$). This is due to a larger fraction of the round time spent in pulling the embeddings for the remote nodes~(Fig.~\ref{subfig:grid-6}) in \textbf{OP} for Products, and thus offering more scope for improvement. Products also is $10\times$ sparser compared to Reddit~(Table~\ref{tab:dataset-specs}). This offers more scope for pull optimizations because a training node is less likely to have many \textit{pull nodes} in its computation graph, which means fewer accesses from the local cache during the forward pass. 
We also observe a steep drop in the median round time for \textbf{OPG} over \textbf{OP} and \textbf{E}~(Fig.~\ref{subfig:grid-2}).

Our strategies also outperform the baseline \textbf{E} for this graph. We see a reduction of $\approx1.9\times$, $\approx3\times$, and $\approx3.6\times$ in TTA for \textbf{OP}, \textbf{OPP}, and \textbf{OPG} strategies over \textbf{E} for GraphConv~(Fig.~\ref{subfig:grid-2}), and comparably for 
SAGEConv. These are consistent with the reduction in median round times over \textbf{E}~(Figs.~\ref{subfig:grid-6} and~\ref{subfig:gs5}) without tangibly increasing the rounds taken, and the convergence plot (Fig.~\ref{subfig:grid-10}), where \textbf{OPG} (pink line) converges the fastest for GraphConv.

However, having a large number of push nodes poses a significant challenge for push optimizations, as they are unable to fully overlap and mask the transfer of embeddings for the vast number of pull nodes within the relatively short training time. E.g., for \textbf{OP} and \textbf{OPG}, the push time~(green stack) exceed the train time for both GraphConv and SAGEConv models~(Figs.~\ref{subfig:grid-6} and~\ref{subfig:gs5}). Since the train time of \textbf{OP} for both models increases due to on-demand embedding pulling, it can hide the push time well.

\subsubsection{Arxiv}

We observe a peak accuracy improvement of approximately $2.1\%$ for GraphConv and $1.4\%$ for SAGEConv when using \textbf{E} and \opes optimizations compared to \textbf{D}. Furthermore, the \textbf{OP} achieve the target accuracy $1.1\times$ and $1.5\times$ faster for GraphConv and SAGEConv, respectively, when compared to \textbf{E}.

However, as the smallest dataset used in this study, Arxiv does not exhibit the same level of benefits from the \opes strategies as the other graphs. It has a significantly smaller subgraph on the clients, and fewer push and pull nodes compared to the other three datasets. Hence, only a small fraction of the total round time is spent in pulling and pushing embeddings in \textbf{E} (Figs.~\ref{subfig:grid-7} and \ref{subfig:gs6}). So \opes has little scope to optimize and is often comparable to \textbf{E}.
The median round times of all \opes strategies being within $< 10\%$ of each other and the accuracy convergence is closely aligned.

In fact, the \textbf{OPP} strategy is penalized due to its optimizations and has the highest round time (Fig.~\ref{subfig:grid-7} and~\ref{subfig:gs6}). Its high volume of pull requests to the embedding server due to a large number of mini-batches~(Arxiv uses a small batch size of $64$) within an epoch add up to a significant overhead~(hatched blue regions in Figs.~\ref{subfig:grid-7} and ~\ref{subfig:gs6}). While \textbf{OPG} has the fastest per-round time, it exhibits a sub-optimal TTA for GraphConv 
due to numerous high-scoring nodes, similar to Reddit.

\subsubsection{Papers}

Since Papers has an extremely high run time of $\approx 20$ hours for $50$ rounds, we only report results on GraphConv model.
\textbf{E} and \opes strategies see an improvement in peak accuracy of $\approx 3.8\%$ over \textbf{D}. Similar to Products, Papers also has a high number of pull nodes for each of the 8 client~($\approx10$M per client), and therefore, spends a majority of the round time in pulling embeddings. Hence, it also benefits substantially from our pull optimizations.

However, since the training time constitutes only a small fraction of the round time in \textbf{E}~(Fig.~\ref{subfig:grid-8}), our push overlap has limited effectiveness. But prune optimizations exhibits a clear benefit, possibly because only a tiny fraction of pull nodes appear during training in a round.

Specifically, \textbf{OP}, \textbf{OPP}, and \textbf{OPG} strategies result in a reduction factor of $\approx2.5\times$, $\approx6\times$, and $\approx11\times$, respectively over the baseline \textbf{E}~(Fig.~\ref{subfig:grid-4}). This matches the reduction in median round times~(Fig.~\ref{subfig:grid-8}) -- $\approx2.4\times$, $\approx5\times$, and $\approx10\times$.
The same is also evident 
from the convergence plots in Fig.~\ref{subfig:grid-12}.

\subsubsection{Summary of Comparison}
\label{subsubsec: discussion}

\textbf{D} has the smallest per-round time since it does no exchange of embeddings and uses only its local subgraph for training. It however has the lowest peak accuracies across graph and GNN models, $2$--$4\%$ below \textbf{E}. But it is particularly poor for Reddit, which has a high edge degree and a large number of remote vertices, leading to  
an accuracy of $\approx65\%$ for GraphConv and $\approx75\%$ for SAGEConv. In contrast, \textbf{E} and \opes achieve $\approx 82\%$ and $\approx90\%$ corresponding accuracies. \textbf{E} and \opes also have comparable peak accuracies, within $1.5\%$.

The median round times reduce significantly for the incremental optimizations of \opes compared to \textbf{E} for all datasets, except Arxiv for GraphConv, for which they are comparable. On average, \textbf{O}, \textbf{P}, \textbf{OP}, \textbf{OPP} and \textbf{OPG} respectively reduce the per-round times over \textbf{E} by $\approx1.2\times$, $\approx1.8\times$, $\approx2\times$, $\approx2.9\times$ and $3.7\times$, across all graphs and models evaluated.
As a consequence of these, the \opes strategies incrementally offer the best time-to-accuracy for all graphs other than for Arxiv, where the potential for optimizations is low due to the small graph size and push/pull time.


\subsection{Analysis of Push Overlap Strategy}
\label{subsec:push-analysis}
We next investigate the behaviour of each of our strategies under different conditions. We start with push overlap optimization and make the following claims.

\claim{Pre-emptive push of embeddings has minimal impact on the accuracy of the trained model.}
\noindent We send stale embeddings from the $\epsilon-1$ epoch to the embedding server while the final epoch of training is running on the client.
This helps hide the cost of copying the embeddings with the training time but can lead to poorer accuracy, compared to EmbC, which pushes the final embeddings after the last epoch of training.
We see that the reduction in peak accuracy due to pushing and using stale embeddings for remote vertices is not noticeable. The drop in peak accuracy ranges between $0.1\%$ for Products to $0.75\%$ for Reddit, with the latter having a high remote vertex count and being more sensitive.
The accuracy lag across rounds is more than offset by having a faster per-round time when using push overlap, which leads to a $10\%$ drop in time to accuracy
for Reddit, Products and Papers when using \textbf{O} over \textbf{E}. So we converge faster.

\claim{Push optimization is more beneficial 
if the training time for an epoch is long enough to mask the push phase and associated overheads
completely.}
\noindent For Reddit, strategy \textbf{O} is able to fully eliminate the time taken for push phase compared to the equivalent \textbf{E}, which does not use this optimization, i.e., the green stack of \textbf{E} ($1.8$s, $2$s) is absent in \textbf{O}, in Figs.~\ref{subfig:grid-5} and \ref{subfig:gs4}.
Similarly, \textbf{OP} masks the $0.8$s and $0.9$s push times of \textbf{P} for GraphConv and SAGEConv completely.
On the contrary, for Products (Figs.~\ref{subfig:grid-6} and \ref{subfig:gs5}) and Papers (Fig.~\ref{subfig:grid-8}), \textbf{O} and \textbf{OP} are unable to completely hide the push time, but still show a reduction of $\approx45\%$ in push time. 

The training time for the last epoch for Reddit is $2.4$s which is much larger than the time to push $\approx 100k$ embeddings of $1.8$s on GraphConv using \textbf{E} and \textbf{O}. So the push time does not contribute to the total round time in \textbf{O}.
In contrast, Products takes $1.7$s for the last training epoch of GraphConv, which is smaller than the $8.5$s it takes to push $\approx 400k$ embeddings, which is much larger than for Reddit. Hence, the push time in excess of the per-epoch training time is only partially masked for Products.
We see a similar contrast for Arxiv, where the training time is larger than the push time ($5.7$s vs. $0.5$s for \textbf{E} on GraphConv) which is able to hide the push fully while Papers has $\approx20M$ embeddings per client taking $295$s to push, which cannot be fully overlapped by the much smaller epoch training time of $7$s.
Even so, we see a substantial drop of $2\times$ and $1.8\times$ in the push time for Products and Papers, respectively.

The number of embeddings pushed affects the push time, and for optimizations like \textbf{P} that reduce this count, the push time is smaller. But these also help reduce the training time due to fewer accesses to the local embedding cache,
and hence the ability to hide does not vary much when we move to \textbf{OP}. 
In general, the benefits of push optimization are sensitive the number of training nodes, the number of batches per epoch, or other factors that affects the training or push time.

\claim{Graphs with a large number of embeddings being pushed concurrently can slow down the training phase time.}
\noindent We see an increase of $16\%$, $14\%$ and $32\%$ in the training time~(blue stack of Figs.~\ref{subfig:grid-5}--\ref{subfig:grid-7}) for Arxiv, Reddit and Products when using push optimization. This is because the concurrent push process competes with the training phase process. The increase in training is substantial for Papers due to $\approx 20M$ embeddings  computed per client during the push phase through a forward phase execution. This 
increases the training time for the final epoch, by $80$s and $35$s in the blue stack for \textbf{O} and \textbf{OP} as compared to \textbf{E} and \textbf{P}, in Fig.~\ref{subfig:grid-8}.  

\claim{Push optimizations are not beneficial if the push phase time is a small fraction of the round time.}
In cases where the number of push nodes is low, the launch overheads for the push process may be non-negligible compared to the actual time to send the embeddings. Here, the push benefits are offset by an increase in the training time, as observed in Arxiv~(Figs.~\ref{subfig:grid-7} and \ref{subfig:gs5}). The push times in \textbf{E} and \textbf{P} is just $\approx11\%$ of the total round time as it sends just $25k$ and $16k$ embeddings, respectively. On applying the push optimizations, \textbf{O} and \textbf{OP} show an increase in train time by $\approx10\%$ offsetting the drop in push time. This leads to an overall increase in TTA by $6\%$ for \textbf{O} relative to \textbf{E}.

\begin{figure}[t]
    \centering
    \subfloat[Reddit]{%
    \includegraphics[width=0.45\textwidth]{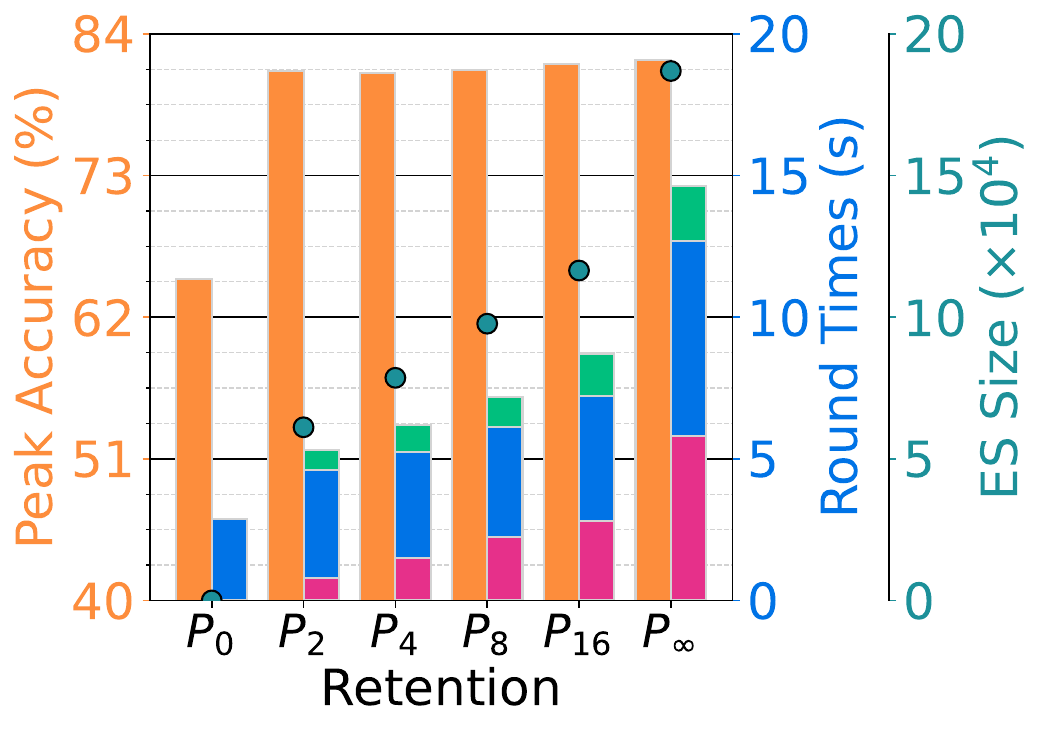}%
    \label{subfig:reddit_pru}%
  }\qquad
  \subfloat[Products]{%
    \includegraphics[width=0.45\textwidth]{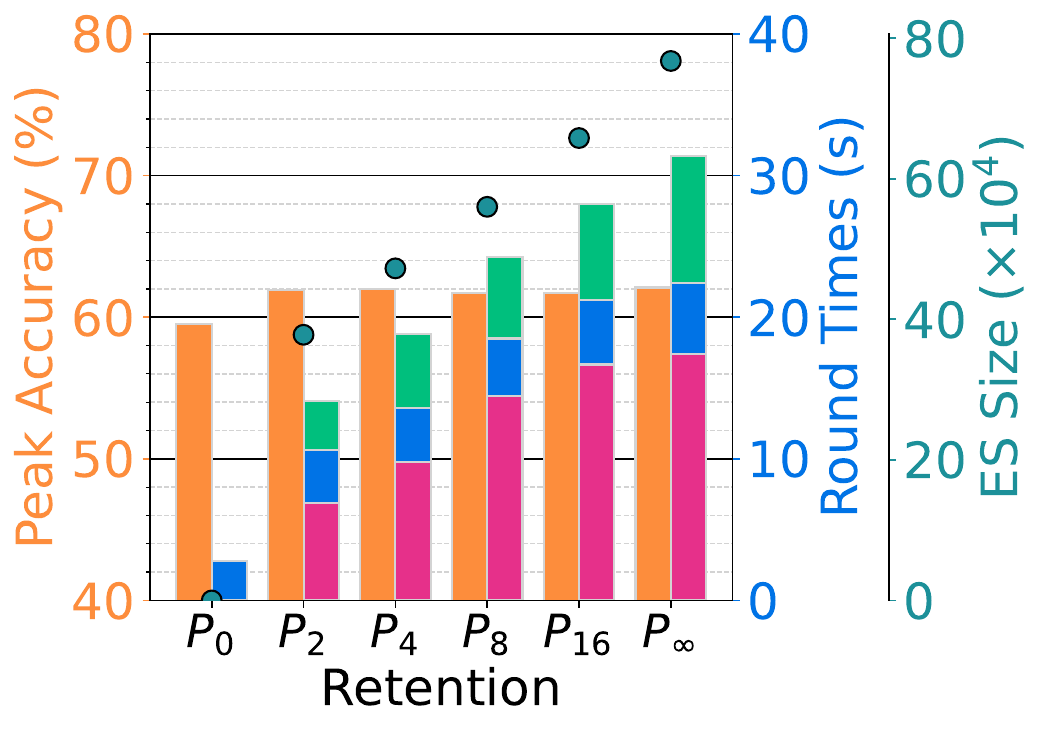}%
    \label{subfig-2:products_pru}%
  }\\
  \subfloat[Arxiv]{%
    \includegraphics[width=0.45\textwidth]{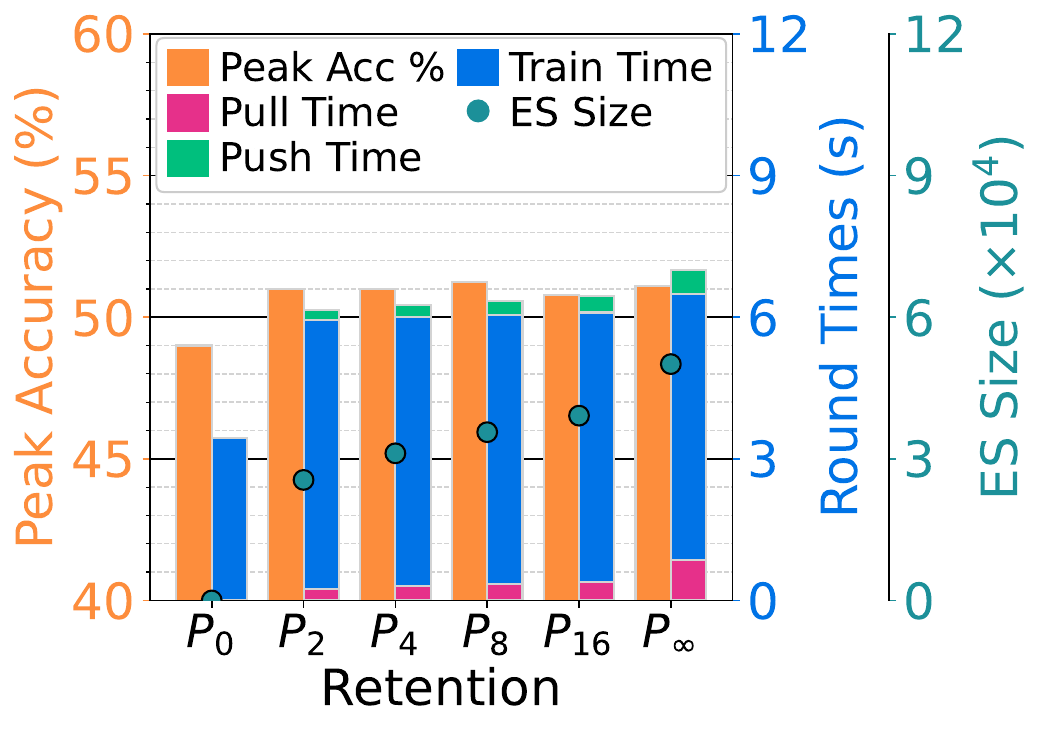}%
    \label{subfig-3:arxiv_pru}%
  }
    \caption{Effect of different numbers of retained nodes~($P_i$) with \textit{uniform random pruning}~(strategy \textbf{P}) on per-round time, peak accuracy achieved and size of embeddings maintained for different graphs.}
    \label{fig:peak-acc-prune}
\end{figure}
\subsection{Analysis of Pruning Strategies}
\label{subsec:prune-analysis}

We explore the effect of using different \textit{retention limits}, with $P_i$ indicating that up to $i$ remote vertices are retained after \textit{uniform random pruning} of the expanded subgraph using \textbf{P}. 
Fig.~\ref{fig:peak-acc-prune} plots the peak accuracy~(orange bar, left Y axis), and the median per-round time~(stacked bar, right inner Y axis) showing the time per phase as well.
We also report the number of embeddings maintained at the embedding server~(marker, right outer Y axis), which correlates with the push and pull times as well as the training time. 
The two extremes are $P_0$, indicating default federated GNN where no remote vertices are used, and $P_\infty$, where all remote vertices are retained and no pruning optimization is done, as done in EmbC.
\begin{figure*}[t!]
    \centering
    \subfloat[Peak acc.~(GC)
    \label{subfig:rev-1.1}]{\includegraphics[width=0.4\textwidth]{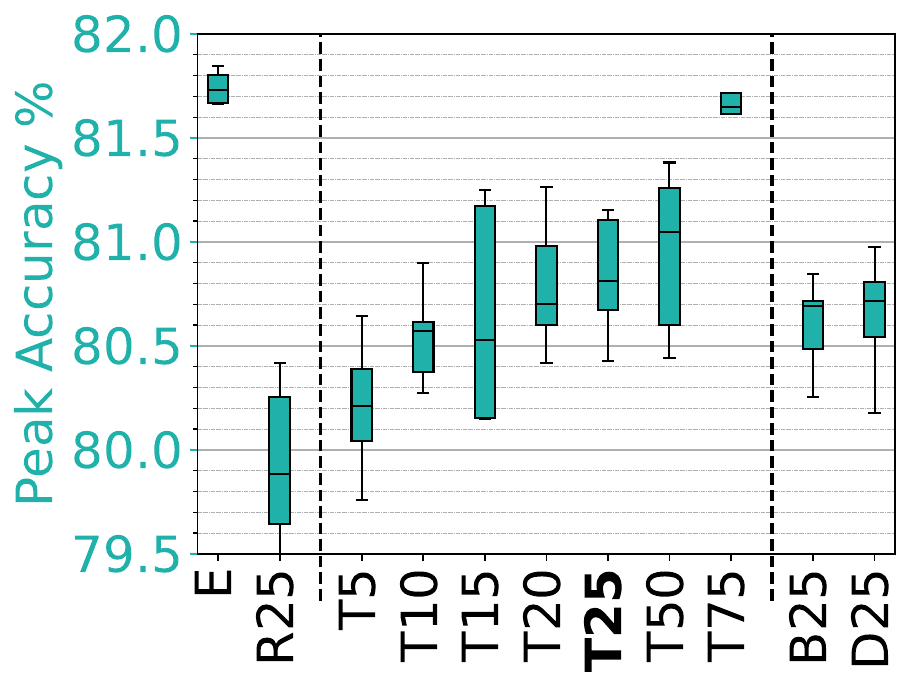}}\hfill
    \subfloat[Time to acc.~(GC)\label{subfig:rev-1.2}]{\includegraphics[width=0.4\textwidth]{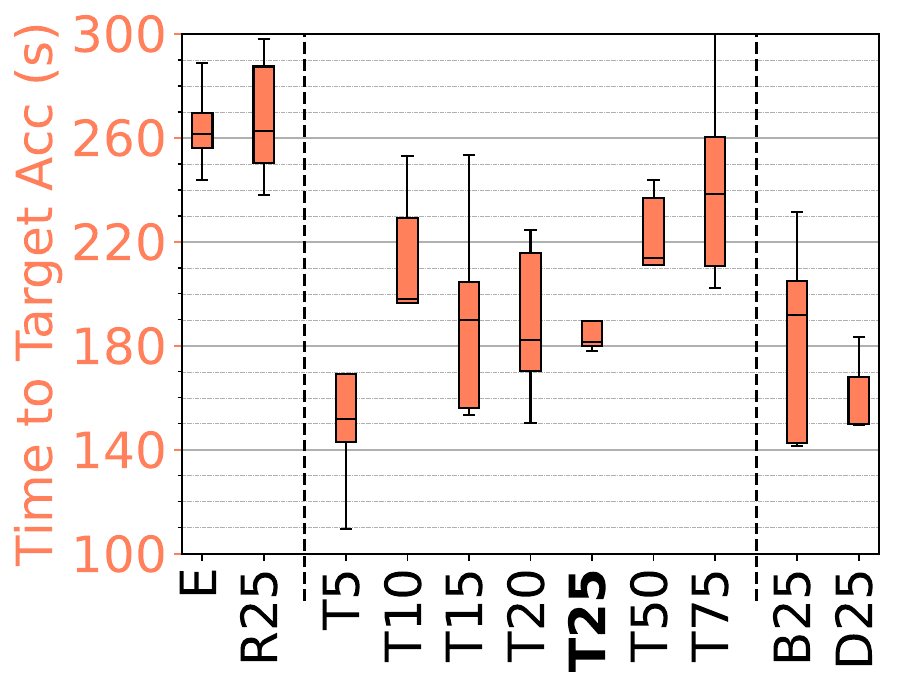}}\\
    \subfloat[Peak acc.~(GS)\label{subfig:rev-2.1}]{\includegraphics[width=0.4\textwidth]{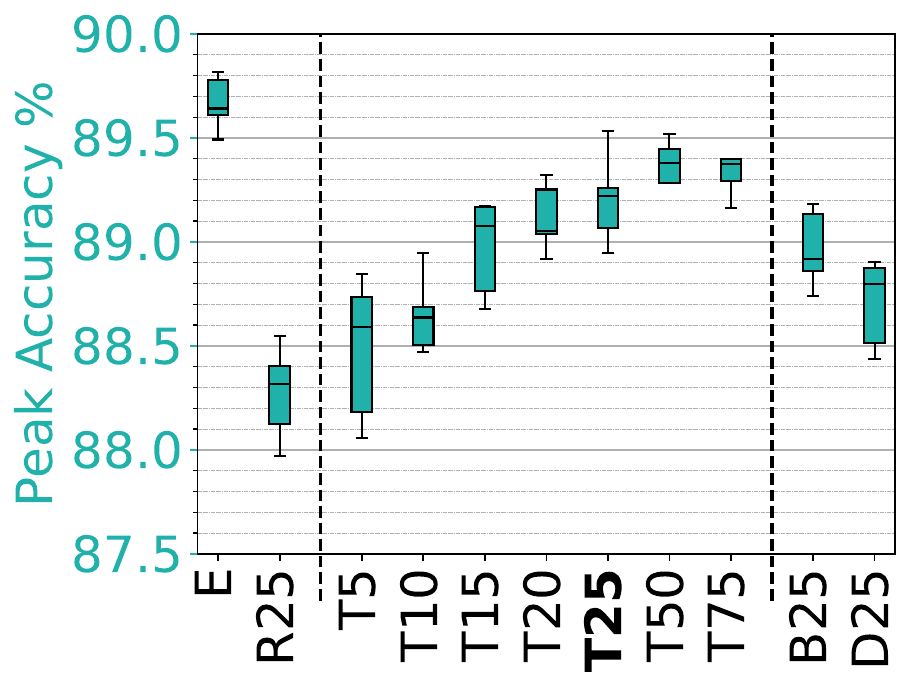}}\hfill
    \subfloat[Time to acc.~(GS)\label{subfig:rev-2.2}]{\includegraphics[width=0.4\textwidth]{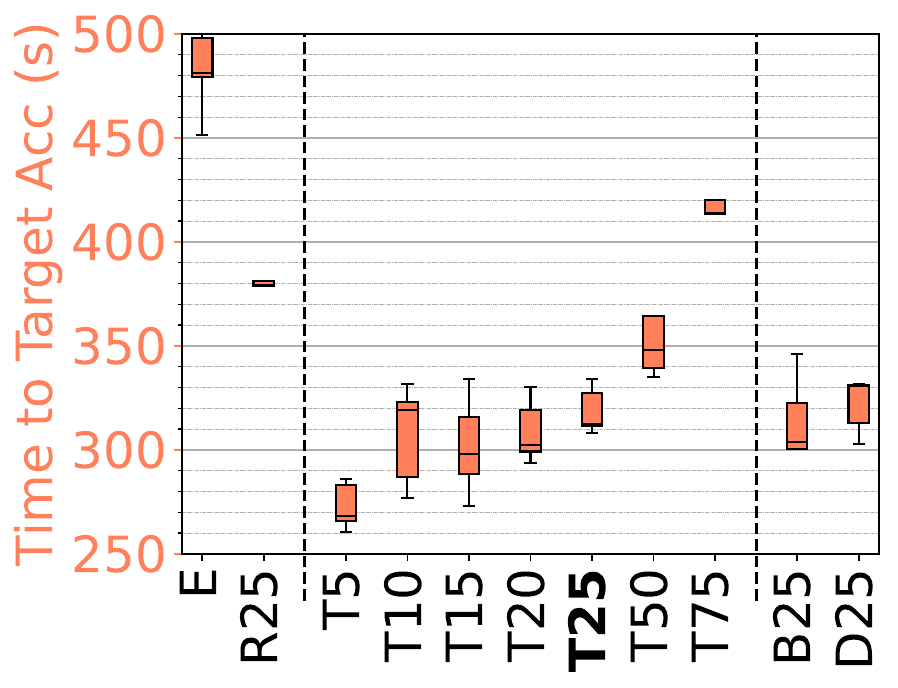}}
    \caption{Ablation study to show the effectiveness of the proposed \textit{frequency-score} scoring strategy~(\textbf{OPG}) for pruning for Reddit dataset using 3-Layer GraphConv~(GC) and SAGEConv~(GS) models. \textbf{E} is the baseline EmbC, \textbf{R25} retains $25\%$ nodes at random, \textbf{T$f$} retains the \textit{top-f\%} nodes by frequency score, and \textbf{B25}/\textbf{D25} retain $25\%$ of nodes based on bridge and degree centrality scores, respectively.}
    \label{fig:revision-plot-1}
\end{figure*}

We further evaluate the effectiveness of the proposed frequency-score pruning strategy~(OPG) through an ablation study across multiple variants. Fig.~\ref{fig:revision-plot-1} reports the Peak Accuracy and Time to Target Accuracy~(TTA) on the Reddit dataset using 3-layer GraphConv and SAGEConv models, respectively. We compare the baseline EmbC without pruning~(E), pruning where $25\%$ of nodes are retained at random~(R25) or based on their frequency score~(T25), other variants where $f\%$ nodes are retained based on their frequency scores~(T$f$), and strategies where $25\%$ of nodes are retained based on bridge~(B25) or degree~(D25) centrality scores.

\claim{Uniform random pruning results in a significant decrease in push and pull time, and a modest reduction in training time, leading to lower per-round time.}
Pruning limits the expanded subgraph size, reducing the number of remote vertices for which embeddings are pulled/pushed in a round~(pink/green stacks of Fig.~\ref{fig:peak-acc-prune}) between the clients and the embedding server.
We see significant reductions in the per-round time for Reddit (Fig.~\ref{subfig:reddit_pru}) and Products (Fig.~\ref{subfig-2:products_pru}), by $2.8\times$ and $2.2\times$, using $P_2$ with the strategy \textbf{P} compared to EmbC's $P_\infty$. 
Retaining just $2$ remote vertices
causes a substantial reduction in the number of embeddings pushed/pulled, reducing them from $226k$ and $568k$ for these two graphs to just $44k$ and $260k$ -- a drop of $5.1\times$ and $2.2\times$. 
This leads to a reduction in their push/pull times per round, e.g, reducing from $9.0$s/$17.4$s for Products without pruning to $3.5$s/$6.8$s with $P_2$. This also drops their per round time by $36.6\%$ and $44.7\%$ for Reddit and Products.

The Reddit and Products graphs are denser and have more remote vertices, leading to higher benefits of pruning.
In contrast, Arxiv has a modest reduction in push/pull times 
from $0.46$s to $0.20$s/$0.79$s to $0.20$s
since the number of embeddings due to pruning reduces from $30k$ to $14k$.
This leads to a $11.8\%$ reduction in per-round time.
Lastly, we observe a slight reduction in training time per round when fewer nodes are retained, likely due to fewer embeddings being loaded from the local cache and a less expensive embedding matrix update during the forward pass.

\claim{Uniform random pruning achieves comparable peak accuracies, and ensures accelerated TTA.}
Fig.~\ref{fig:peak-acc-prune} shows a significant reduction in round times, as the retention limits are decreased from $P_\infty$ to $P_2$, with a minimal loss of accuracy. E.g., for Products, the peak accuracy remains at between 61.8\%--62.1\% while the time decreases from 31.6s--14.1s when we go from $P_\infty$ to $P_2$.
This follows from our intuition in \S~\ref{subsec:prune} that not all remote nodes add to the model performance. As expected, the peak accuracy shows a significant drop for default federated GNN ($P_0$), by $16\%$ for Reddit and $2.2\%$ for Products. 
Comparable peak accuracies combined with a consistent decrease in round time as retention limit decreases ensures a faster TTA for lower retention limits as compared to $P_\infty$ or \textbf{E}.

\claim{Our scoring strategy enables retention of important nodes in scored pruning.}
Pruning reduces the time taken per round, as we see above. However, rather than randomly removing remote vertices, our score-based pruning retains those nodes that are expected to contribute more and potentially improve the accuracy. When we prune $75\%$ of remote vertices at random~(\textbf{OPG}$_{R25}$) or based on the scoring~(\textbf{OPG}$_{T25}$) for GraphConv, the reduction in time taken per round for Reddit relative to EmbC baseline is comparable, both at $\approx3.2\times$. However, random pruning leads to a drop in peak accuracy (Fig.~\ref{subfig:rev-1.1}), from $81.73\%$ for \textbf{E} to $79.88\%$, while our score-based pruning achieves $80.8\%$. This also improves the TTA for scored-pruning, converging $31\%$ faster than its random counterpart (Fig.~\ref{subfig:rev-1.2}). We see similar results for SAGEConv where \textbf{OPG$_{T25}$} at $89.22\%$ peak accuracy outperforms \textbf{OPG$_{R25}$} with $88.32\%$ peak accuracy with $\approx18\%$ improvement in TTA~(Figs.~\ref{subfig:rev-2.1} and~\ref{subfig:rev-2.2}).

\claim{Threshold of $f=25\%$ provides a reasonable trade-off between peak accuracy and TTA.}
Fig.~\ref{fig:revision-plot-1} shows a more detailed ablation study demonstrating how different values of $f$ affect performance on the Reddit dataset using 3-layer GraphConv and SAGEConv models. We see a clear trend in the peak accuracy~(Fig.~\ref{subfig:rev-1.1}) as we move from $f=5\%$ to $f=75\%$, with a peak accuracy difference of up to $2.3\%$ between \textbf{OPG}$_{T5}$ and \textbf{OPG}$_{T75}$. 
We notice similar trends in peak accuracy for a 3-layer SAGEConv~(Fig.~\ref{subfig:rev-2.1}) as well, where this peak difference is $1.8\%$.
However, this increase in peak accuracy comes at the cost of increased round times, leading to higher time to target accuracies, with the median TTA increasing by up to $\approx1.6\times$ from \textbf{OPG}$_{T5}$ to \textbf{OPG}$_{T75}$~(Fig.~\ref{subfig:rev-1.2}). 
We chose $25\%$~(\textbf{OPG$_{T25}$}) for our main experiments because it provides a good balance between the highest accuracy achieved and the time it takes to reach the target accuracy.

\claim{Frequency-score marginally outperforms centrality-based scoring measures.}
We evaluate the effectiveness of the frequency score in comparison to other metrics such as degree centrality and bridge centrality. Degree centrality, defined by the number of neighbours a node possesses, serves as a measure of a node's local connectivity or popularity. In contrast, bridge centrality captures a node's role in facilitating connectivity between distinct communities or subgraphs within the graph. Both metrics are relevant in the context of GNN \textit{message passing}. Nodes with high degree centrality can influence a large number of neighbours, whereas nodes with high bridge centrality imply their ability to relay information across communities in a graph, contributing to global information flow during training.

Figs.~\ref{subfig:rev-1.1} and~\ref{subfig:rev-2.1} show that \textbf{OPG$_{T25}$} marginally outperforms both its bridge and degree centrality counterparts~(\textbf{OPG$_{B25}$} and \textbf{OPG$_{D25}$}, respectively) across both models. This is because while centrality measures capture the structural importance of a node, they do not necessarily reflect how often that node is actually involved in message passing during training (i.e., how many training nodes use it in a given epoch). However, we do not see improvements in terms of time-to-accuracy~(Figs.~\ref{subfig:rev-1.2} and~\ref{subfig:rev-2.2}). This is because of increased round times for \textbf{OPG$_{T25}$} due to a higher number of \textit{push nodes} per client as compared to the other two strategies. In particular, \textbf{OPG$_{T25}$} pushes embeddings for an average of $\approx20\%$ more nodes per client in every round, thus increasing round times. Centrality-based metrics select pull nodes based on their local subgraph importance, leading to high overlap in pull sets across clients. As a result, each client has fewer unique nodes to push embeddings for. In contrast, our frequency score metric selects nodes based on actual training usage, resulting in more diverse pull sets and a higher push load per client.

Lastly, we explain the rationale behind picking the frequency score over other metrics. The frequency score measures how often a remote node is used by local training nodes during message passing. This makes it a good indicator of how useful a node is during training, both in terms of structure and actual usage across different batches. Nodes accessed more frequently are likely to contribute to more gradient updates across the training process. While this score doesn’t directly capture task-specific gradients or semantic meaning, it is a reasonable and lightweight proxy for
identifying nodes that are repeatedly useful during training. Importantly, it does not compromise privacy in a federated setting, as it doesn’t require sharing node features or neighborhood information.

\claim{Score-based pruning reduces the round times but does not provide the best time-to-accuracy for dense graphs.}
\noindent We notice a sharp drop in round times for score-based pruning~(\textbf{OPG}) over \textbf{E}, providing a reduction of $\approx3.6\times$ averaged across all datasets and across both models~(last bar, Figs.~\ref{subfig:grid-5}--\ref{subfig:grid-8} and~\ref{subfig:gs4}--\ref{subfig:gs6}). However, this does not always translate to better TTA for all datasets. E.g., for Reddit, even with a significant drop in round time for both models, we see an increase in TTA for GraphConv while the TTA for SAGEConv remains the same for \textbf{OPG} over \textbf{OPP}. Reddit, being a dense graph, loses many critical high-scoring remote nodes when only $25\%$ nodes are retained, which leads to information loss, thus resulting in a longer time to accuracy~(more rounds to convergence).

\begin{figure*}[t]
\centering
    \subfloat[Nodes pulled per RPC when training in \textbf{OPP}$_{T0}$, \textbf{OPP}$_{T25}$ and \textbf{OPP}$_{R25}$.\label{subfig:pull-analysis-1}]{\includegraphics[width=0.37\textwidth]{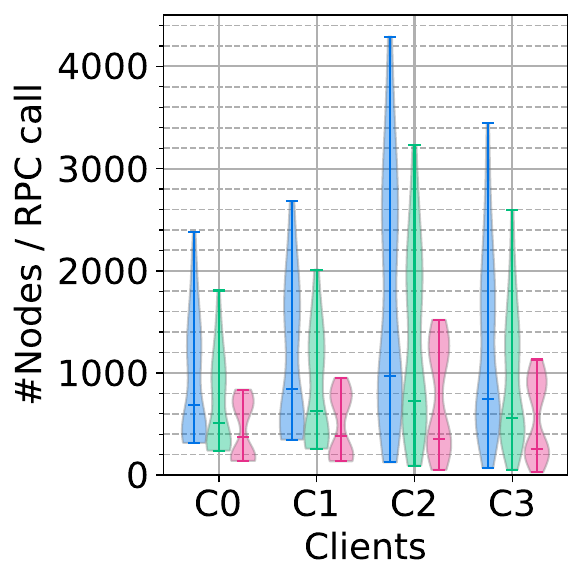}} \qquad
    \subfloat[RPC request time during training in \textbf{OPP}$_{T0}$, \textbf{OPP}$_{T25}$ and \textbf{OPP}$_{R25}$.\label{subfig:pull-analysis-2}]{\includegraphics[width=0.37\textwidth]{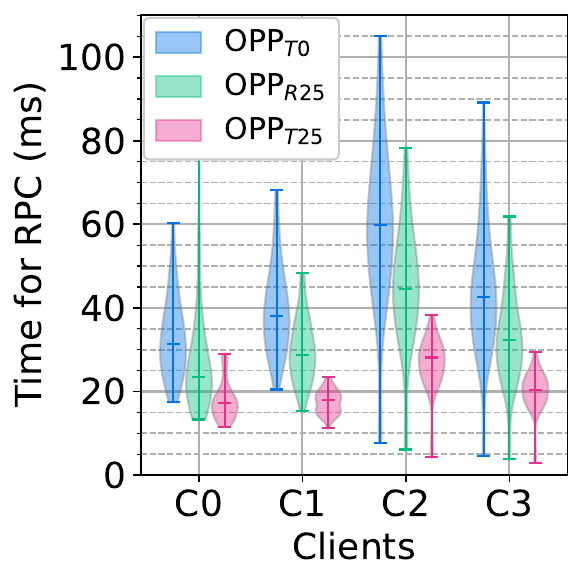}} \\
    \subfloat[Correlation between the number of nodes per request and the time to service that request.\label{subfig:pull-analysis-3}]{\includegraphics[width=0.37\textwidth]{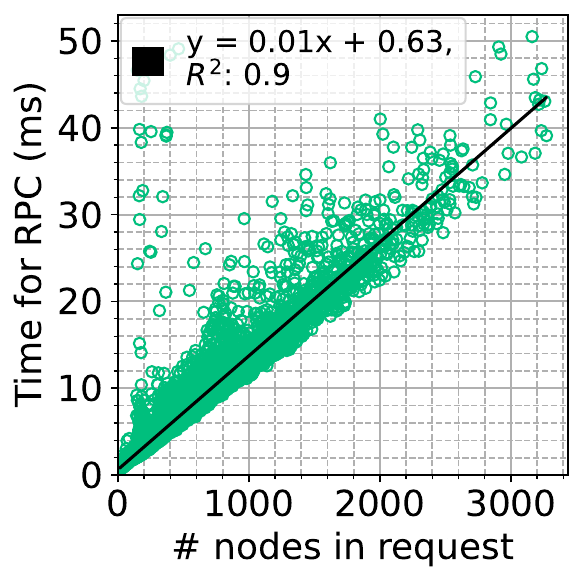}}\qquad
    \subfloat[The total pull time~(initial pull time in pink stack + time spent pulling during training in hatched blue stack) for different batch sizes.\label{subfig:pull-analysis-4}]{\includegraphics[width=0.37\textwidth]{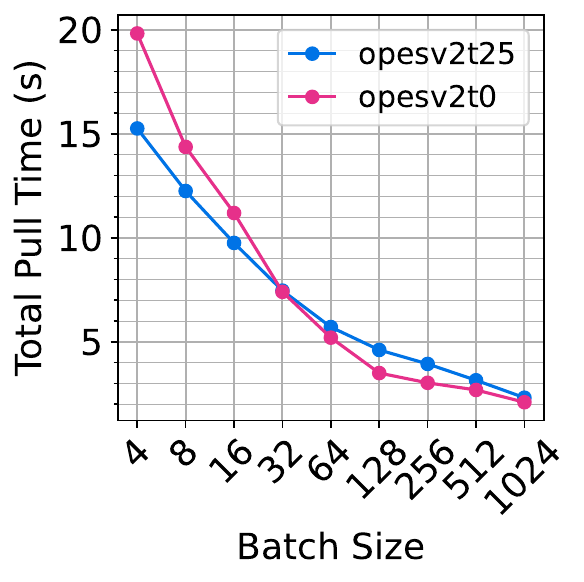}}\\
    \caption{Plots for analysis of strategies for pull phase optimization. All results reported for Products.}
    \label{fig:pull-opt-analysis}
\end{figure*}

\subsection{Analysis of Pull Phase Pre-fetch Strategy}
\label{subsec:pull-analysis}
Fig.~\ref{fig:pull-opt-analysis} demonstrates the analysis of the scored pre-fetch pull strategy. Figs.~\ref{subfig:pull-analysis-1} and~\ref{subfig:pull-analysis-2} show the distribution of number of nodes per call to the embedding server and the time to service those calls, respectively, while Fig.~\ref{subfig:pull-analysis-3} correlates the two, presenting a good fit~($R^2=0.9$). Lastly, we show how the total pull time varies with the number of minibatches~(smaller minibatches sizes correspond to a larger number of minibatches).

\claim{Pull phase pre-fetch is beneficial when the pull phase forms a significant fraction of the round time but suffers for dense graphs.}
Products and Papers spend a large part of their round time on the pull phase (pink stack of Figs.~\ref{subfig:grid-6},~\ref{subfig:grid-8}). When we include pull phase pre-fetch (\textbf{OPP}) over \textbf{OP}, we see a reduction in the pull phase times by $7.5$s and $280$s for these two graphs because embeddings for fewer nodes are pulled overall in the round.
E.g., \textbf{OPP} pre-fetches embeddings for $25\%$ of high-scoring nodes upfront for Products, which  suffices to meet almost half of the requirements, and additional embeddings for only $\approx25\%$ of pull nodes are pulled dynamically over $32$ RPC calls, compared to pulling embeddings 
for all $100\%$ of pull nodes beforehand by \textbf{OP}.
This leads to lower round times by $3.7\times$ and $5.1\times$ for \textbf{OPP}
and also faster TTA by $3.1\times$ and $6.9\times$~(Fig.~\ref{subfig:grid-2},~\ref{subfig:grid-4}).

However, 
for a dense graph like Reddit, most pull nodes have a high score. So, pre-fetching embeddings for only $25\%$ of the top-scoring nodes in the pull phase still leaves out other high-scoring ones, which results in an additional $\approx40\%$ being dynamically pulled during training. As a result, Fig.~\ref{subfig:grid-5} shows a reduction in pull phase time for \textbf{OPP} relative to \textbf{OP} by $1.2$s (pink stack) but this is offset by an increase in the dynamic pull time by $0.6$s (hatched blue stack), resulting in just a modest reduction in the round time by $10\%$.

For Arxiv, \textbf{OPP} dynamically pulls embeddings for an additional $37\%$ remote nodes beyond $25\%$ pulled initially over $\approx500$ dynamic pull calls, with a small number of nodes per call, leading to significant RPC call overheads.
Since Arxiv uses a smaller batch size of 64, it has a much larger number of minibatches compared to other datasets. The overhead of numerous small network calls effectively nullifies the benefits of \textbf{OPP}.

\claim{Our scoring strategy enables fewer embeddings to be pulled in each RPC call during training, leading to a lower time to service the call.}
By default, \textbf{OPP} pre-fetches embeddings for $25\%$ of high scoring nodes (\textbf{OPP}$_{T25}$). We compared it against pre-fetching a random $25\%$ of pull nodes (\textbf{OPP}$_{R25}$) for Products. 
\textbf{OPP}$_{T25}$ dynamically pulls $45\%$ fewer embeddings per call than \textbf{OPP}$_{R25}$, leading to a $38\%$ drop in RPC call time (Figs.~\ref{subfig:pull-analysis-1} and~\ref{subfig:pull-analysis-2}, green violins versus red violins).
This reduction is due to the initial score-based pre-fetch that is able to effectively identify nodes that will more often be used during training, resulting in far fewer being pulled on-demand.

The number of nodes per RPC call can be correlated to the time to service each call as shown in Fig.~\ref{subfig:pull-analysis-3}, which, along with Fig.~\ref{subfig:pull-analysis-4} demonstrates that the time-to-service the calls for dynamic pulls also reduces for the scored pre-fetch. 

\begin{figure}[t]
\centering
    \centering
    \subfloat[Reddit\label{subfig:scale-parts-reddit}]{%
          \includegraphics[width=0.5\textwidth]{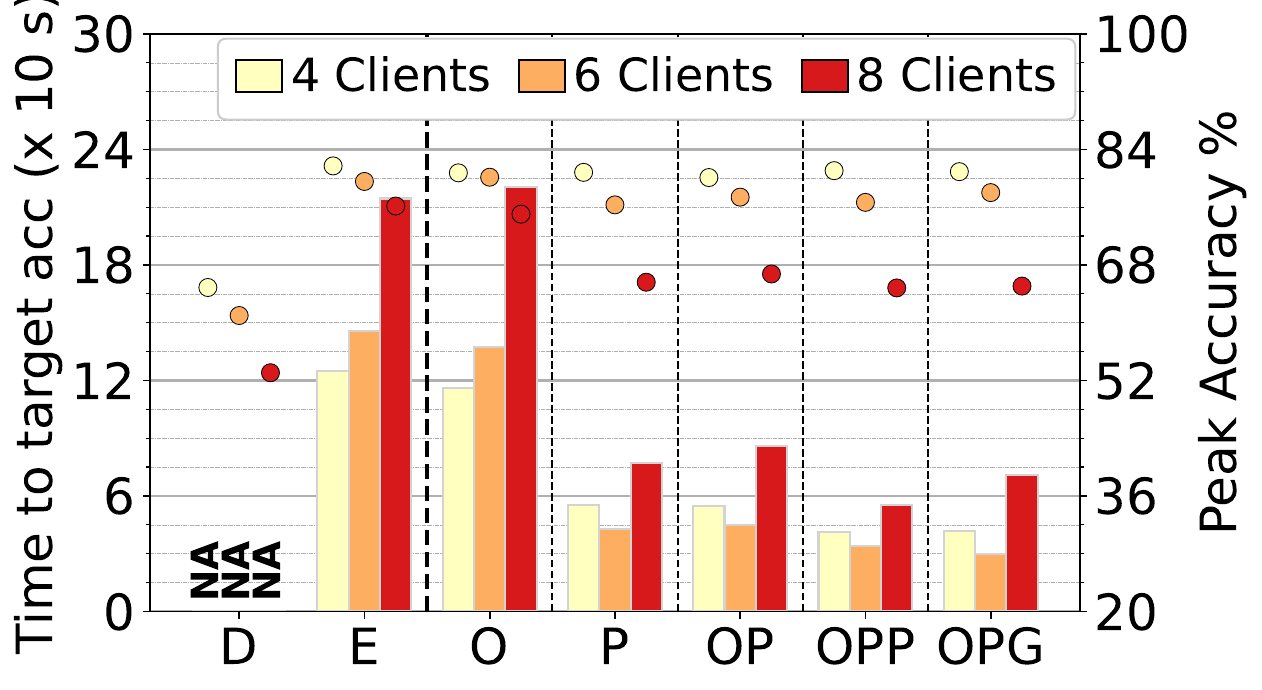}
    }
    \subfloat[Products\label{subfig:scale-parts-products}]{%
          \includegraphics[width=0.5\textwidth]{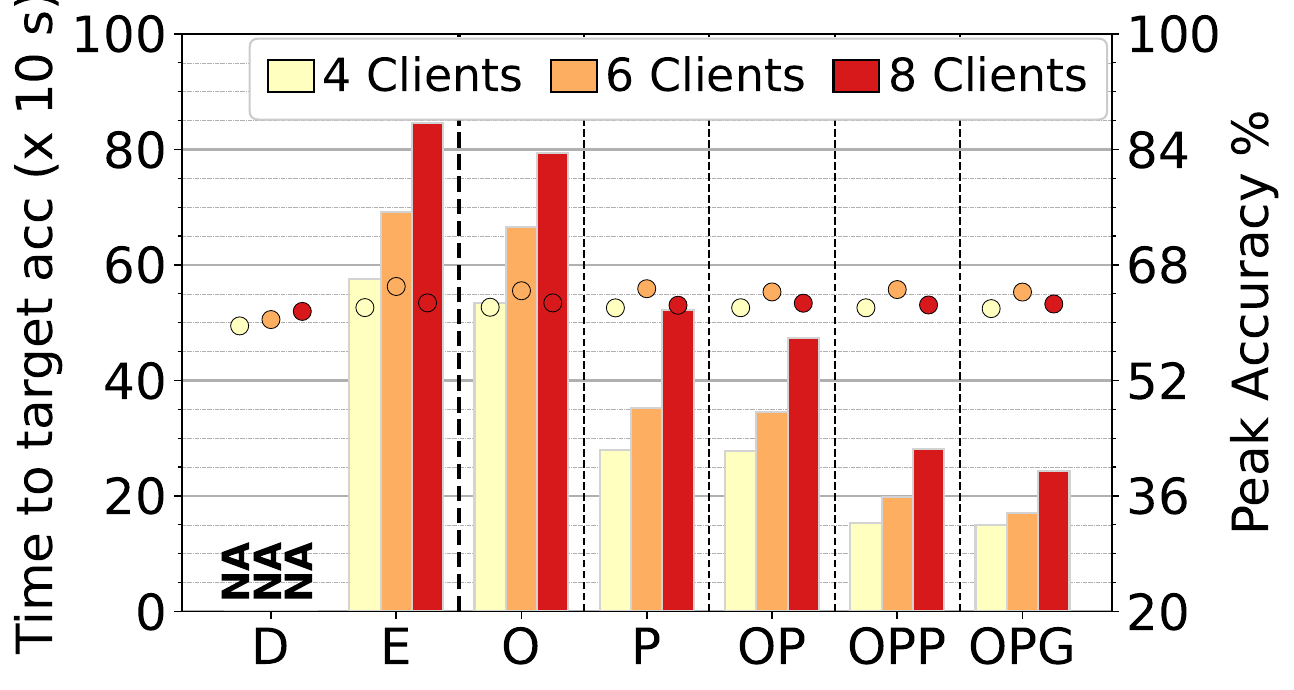}
        }    
    \caption{Time-to-accuracy (left Y axis, bar) and Peak accuracy (right Y, marker) for strategies, with $4,6$ and $8$ clients.}
    \label{subfig:scale-parts}
\end{figure}

\begin{figure}[t]
\centering
    \includegraphics[width=0.5\textwidth]{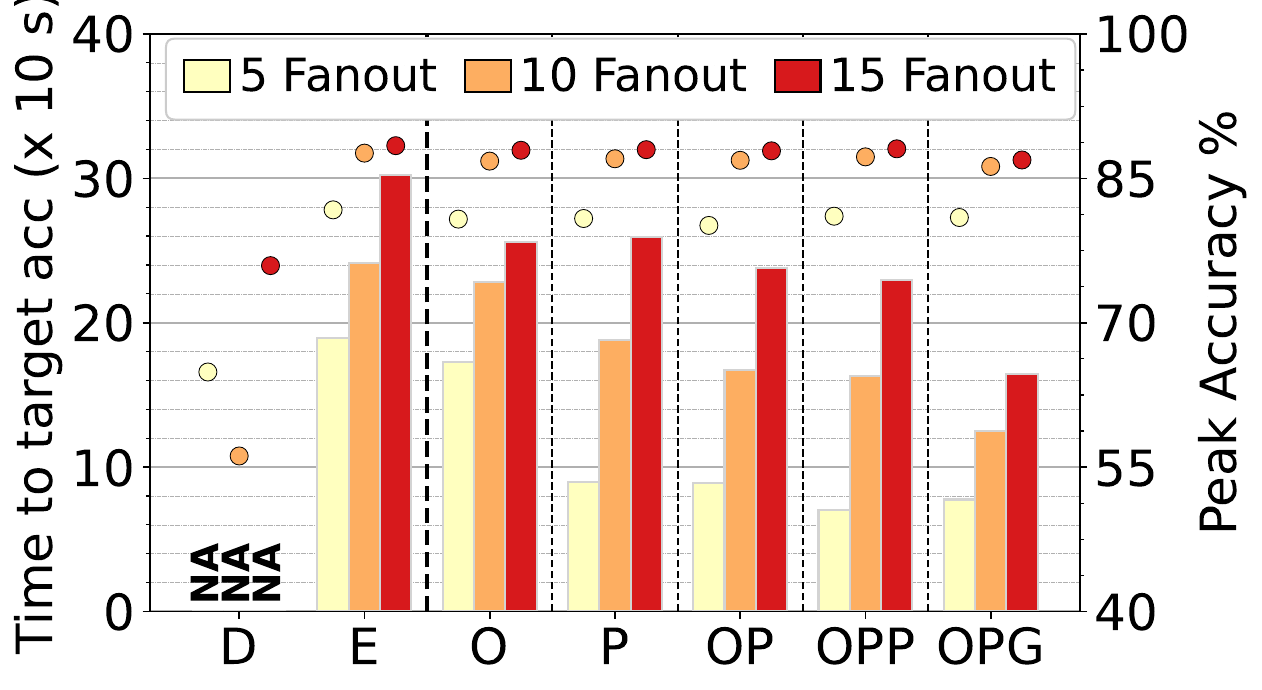}
    \caption{Time-to-accuracy and peak accuracy plots with increasing fanout.}
    \label{subfig:scale-fanout-reddit}
\end{figure}

\claim{Pull pre-fetch benefit depends on the \% of initially pulled nodes and the number of calls to the embedding server during dynamic pull. 
}
We compare two configurations - the default case \textbf{OPP}$_{T25}$, and an extreme case where all nodes are pulled during training \textbf{OPP}$_{T0}$~($0\%$ nodes pulled initially). Either approach offers unique advantages depending on the number of minibatches and pull requests needed during training.

For example, in a setting where each round consists of one epoch with \textit{N} minibatches, \textbf{OPP}$_{T0}$ makes maximum \textit{N} requests to the embedding server, with each request retrieving a larger set of vertices per call as compared to \textbf{OPP}$_{T25}$~(blue violin vs. red violin in Fig.~\ref{subfig:pull-analysis-1}). Therefore, while \textbf{OPP}$_{T0}$ may have a higher service time per call due to the larger volume of data being handled, it benefits from fewer overall calls. For scenarios involving a smaller number of batches, it can often perform more efficiently because the overhead of making fewer calls outweighs the higher service time per call.

In contrast, \textbf{OPP}$_{T25}$ makes $N + 1$ calls, with the first call handling $25\%$ of the nodes and the subsequent $N$ calls retrieving much fewer nodes per call. Each call processes a smaller batch of data~(Fig.~\ref{subfig:pull-analysis-1}), leading to a reduction in the service time per call~(Fig.~\ref{subfig:pull-analysis-2}). This strategy can be advantageous when dealing with a larger number of minibatches (i.e., a higher number of pull requests), where the time required for each of \textbf{OPP}$_{T0}$'s larger calls can accumulate significantly, making it inefficient~(as observed in Fig.~\ref{subfig:pull-analysis-4}).

\subsection{Scaling Performance with the Number of Clients}

We evaluate the scaling performance of \opes as the number of clients increases from $4$ to $6$ to $8$, for Reddit and Products.
Figs.~\ref{subfig:scale-parts-reddit} and~\ref{subfig:scale-parts-products} show the TTA~(bar, left y-axis) and the peak accuracy~(marker, right y-axis) for the various strategies for the GraphConv model.

\claim{The benefits of our \opes strategies are retained even as the number of clients increases.} The \opes strategies maintain their performance trends for all three client counts for both graphs. As the strategies become more complex, they offer incrementally better benefits. While \textbf{O} is marginally better than the baseline \textbf{E}, \textbf{OPP} or \textbf{OPG} are the best strategies in all cases, as seen earlier with $4$ clients (Figs.~\ref{subfig:grid-1} and~\ref{subfig:grid-2}).

\claim{The time to accuracy increases with an increase in the number of clients a graph is partitioned onto.}
Across the three client cohort sizes, the time to target accuracy generally increases as the number of clients rises for both Reddit and 
Products. This is because, the number of training nodes per client reduces as the number of clients increases, which means that the local model receives fewer updates during a round.

One exception to this is Reddit with $6$ clients. The overlap-based strategies show that with $6$ clients, the training time is sufficient to efficiently mask the push phase. However, with $4$ clients, the training time is too long, and with $8$ clients, the push phase is not effectively hidden~(round times confirm this).

Therefore, while the round time decreases as the number of clients increases, more rounds are required to reach the target accuracy. In a practical federated learning setting, however, increasing the number of clients would not reduce subgraph sizes. Instead, the new clients' subgraphs would introduce more information, potentially improving time to accuracy.

\claim{Peak accuracies degrade for \opes for dense graphs with an increase in the number of clients.}
For Reddit, we observe that the drop in accuracy due to pruning methods becomes more significant as the number of clients increases~(Fig.~\ref{subfig:scale-parts-reddit}). The size of the resident subgraph decreases as the graph is split into a larger number of partitions, thereby increasing the fraction of nodes affected due to pruning. This translates to more loss of information. However, we do not observe a significant accuracy drop for the Products dataset. This is because Products already show only a marginal accuracy improvement with embedding-sharing over \textbf{D}, so the exclusion of remote vertices does not greatly affect the peak accuracy. As noted earlier, in a practical scenario, subgraph sizes wouldn't decrease with an increase in the number of clients, so a reduction in peak accuracy shouldn't occur.

\subsection{Effect of Hyperparameters on \opes Strategies}
\label{subsec:fanout}
Lastly, we investigate the effect of hyperparameters and model configurations
on the \opes strategies.

\claim{The time to accuracy increases with an increase in fanout, but the benefits of our \opes strategies are retained.}
Fanout for a GNN training is the number of neighbouring nodes sampled from each node while building the computation graph. These can include both local and remote nodes. 
Fig.~\ref{subfig:scale-fanout-reddit} illustrates the impact of increasing the fanout for the Reddit dataset.
As expected, the time spent in the training phase grows because larger fanouts lead to larger computation graphs. Consequently, we observe an increase in time-to-accuracy for all \opes optimizations as fanout values rise.

\claim{Increase in fanout leads to a better peak accuracy but plateaus to a stable accuracy.}
Fig.~\ref{subfig:scale-fanout-reddit} illustrates that increasing the fanout from $5$ to $10$ leads to nearly a $5\%$ boost in peak accuracy, but further improvement is modest when increasing the fanout from $10$ to $15$, increasing accuracy only by $\approx1.5\%$. This is likely due to the model performing aggregating over a larger neighbourhood as the fanout grows, which may include irrelevant or less useful nodes and can lead to over-smoothing~\cite{liu2020deepergnn}.

\claim{For larger fanout sizes, the TTA of \textbf{OPP}, which is generally lower than \textbf{OPG} for Reddit, reverses, with \textbf{OPG} exhibiting a faster TTA.}
For a fanout of $5$, \textbf{OPP} achieves the shortest time to target accuracy. However, this trend reverses at higher fanouts of 10 and 15. The reason is that as the computation graph expands with fanouts, \textbf{OPP} pulls a greater number of embeddings during training, which increases the overall round time. For the same reason, we also notice a much sharper rise in \textbf{OPP} with fanout values as compared to other optimizations. With a similar peak accuracy, \textbf{OPG} becomes a more favourable choice at larger fanouts.

We separately study the impact of the number of layers of a GNN model, for values of $3$, $4$, and $5$, where we see the \textbf{OPG} strategy outperforming others while \textbf{OPP} deteriorates because of the larger number of RPC calls for embeddings at multiple hops. Lastly, we do not see any interesting trends when we vary the hidden dimension sizes from the default $32$ to values of $64$, $128$, and $256$.

\section{Conclusion}
\label{sec:conclude}
In this article, we presented Optimized Embedding Server (\opes), an optimized approach to the SOTA embedding-based federated GNN learning over subgraphs, like EmbC. \opes optimizations reduce the significant communications required in each federated learning round while maintaining accuracies on par with the SOTA. We propose three high-level optimizations: overlapping embedding communication with training, pruning subgraph expansion, and dynamically pre-fetching pull embeddings. Additionally, we propose and analyze a scoring strategy to further enhance the mentioned optimizations. These help the embedding server scale with increased GNN depth and embedding size.

Our detailed experiments with 4 graphs and 2 GNN models successfully show consistent improvements in time-to-accuracy and per-round time metrics compared to the default federated GNN and EmbC SOTA baselines for the progressively more sophisticated strategies of \opes. We also perform a rigorous analysis of the factors affecting the same. Our brief ablation study also highlights the effects of other features, such as fanout and the number of FL clients, and their impact on the peak accuracy and TTA. As future work, we intend to extend this study to incorporate support for handling data heterogeneity in federated graph learning. Additionally, we also plan to take into account the variability of load on selected clients and propose optimizations for load balancing.
Lastly, we also plan to explore richer scoring mechanisms based on task relevance and data heterogeneity to come up with improved optimizations for pruning.

\bibliographystyle{plain}
\bibliography{arxiv-r1/references.r1}
\end{document}